\documentclass{article}

\usepackage{arxiv}

\usepackage[utf8]{inputenc}
\usepackage[T1]{fontenc}

\usepackage{graphicx}
\usepackage[export]{adjustbox}
\usepackage{subcaption}

\usepackage{array} 
\usepackage{arydshln}
\usepackage[ruled,vlined]{algorithm2e}
\usepackage{mathtools}
\usepackage{multirow}
\usepackage{booktabs}
\usepackage[font=small]{caption}
\usepackage{threeparttable}

\newcommand{\subfigimg}[3][,]{%
  \setbox1=\hbox{\includegraphics[#1]{#3}}
  \leavevmode\rlap{\usebox1}
  \rlap{\hspace*{0pt}\raisebox{\dimexpr\ht1-.2\baselineskip}{{\footnotesize \usefont{T1}{DejaVuSans-TLF}{b} h #2}}}
  \phantom{\usebox1}
}
\newcommand{\beginsupplement}{%
        \setcounter{table}{0}
        \renewcommand{\thetable}{S\arabic{table}}%
        \setcounter{figure}{0}
        \renewcommand{\thefigure}{S\arabic{figure}}%
        \setcounter{equation}{0}
        \renewcommand{\theequation}{S\arabic{equation}}%
        \setcounter{section}{0}
        \renewcommand{\thesection}{S\arabic{section}}%
     }
\DeclareMathOperator{\Tr}{Tr}

\usepackage{lipsum}

\title{Feedback from tissue mechanics self-organizes efficient outgrowth of plant organ}
\author{
  Jason Khadka\textsuperscript{1}, Jean-Daniel Julien\textsuperscript{1}, Karen Alim\textsuperscript{1,2,+} \\
  \textsuperscript{1}Max Planck Institute for Dynamics and Self-Organization (MPIDS), 37077 G\"{o}ttingen, Germany\\
  \textsuperscript{2}Physik-Department, Technische Universit\"{a}t M\"{u}nchen, Garching, Germany\\
  \textsuperscript{+}Correspondence: \texttt{karen.alim@ds.mpg.de} \\
}

\begin{document}
\maketitle

\begin{abstract}
Plant organ outgrowth superficially appears like the continuous mechanical deformation of a sheet of cells. Yet, how precisely cells as individual mechanical entities can act to morph a tissue reliably and efficiently into three dimensions  during outgrowth is still puzzling especially when cells are tightly connected as in plant tissue.  In plants, the mechanics of cells within a tissue is particularly well defined as individual cell growth is essentially the mechanical yielding of cell-wall in response to internal turgor pressure. Cell wall stiffness is controlled by biological signalling and, hence, cell growth is observed to respond to mechanical stresses building up within a tissue. What is the role of the mechanical feedback during morphing of tissue in three dimensions? Here, we develop a three dimensional vertex model to investigate tissue mechanics at the onset of organ outgrowth at the tip of a plant shoot.  We find that organ height is primarily governed by the ratio of growth rates of faster growing cells initiating the organ to slower growing cells surrounding them. Remarkably, the outgrowth rate is higher when cells growth responds to the tissue-wide mechanical stresses. Our quantitative analysis of simulation data shows that tissue mechanical feedback on cell growth can act via twofold mechanism. First, the feedback guides patterns of cellular growth. Second, the feedback modifies the stress patterns on the cells, consequently amplifying and propagating growth anisotropies. This mechanism may allow plants to grow organs efficiently out of the meristem by reorganizing the cellular growth rather than inflating growth rates.
\end{abstract}

\section*{Statement of Significance}
All areal organs in plants begin as outgrowth from the shoot apical meristem (SAM). Organs are initiated by a rapidly expanding patch of cells on the SAM surface. Yet, it is unclear how quicker cell growth can generate outgrowth, given that cells are tightly connected by shared cell walls within the tissue. Here, we build a three-dimensional vertex model of tissue growth. In particular, we account for mechanical feedback of tissue-wide stresses on cell growth. We find that the mechanical feedback is pivotal for efficient outgrowth as it self-organizes anisotropic growth of outgrowth boundary cells allowing the primordia to bulge out. This mechanism allows for self-organized differentiation of cell growth patterns - likely relevant well beyond the model system studied here.

\section*{Introduction}
Stochastic cellular growth and division result in robust and reproducible shaped tissues and organisms. What leads to this robustness on the tissue-wide scale, despite the apparent stochasticity on the cell scale, has been a puzzling question in biology \cite{hong2016variable,farhadifar2007influence,osterfield2017epithelial,hamant2008developmental}. In plants, cells are enclosed by rigid cell walls and the mechanics of these walls dictates cell growth. Anisotropic stiffness of the walls lead to anisotropic growth of cells \cite{baskin2005anisotropic}. Most strikingly, the growth of cells is coupled mechanically through shared walls. Expansion of one cell is communicated to all immediate neighbors through forces on cell walls and junctions. This mechanical coupling along with biochemical signaling has been proposed as the organizer of growth \cite{heisler2010alignment,nakayama2012mechanical,long2019emergence}. However, a theoretical framework for studying the role of mechanics in dynamically morphing a tissue in three dimensions is still missing to elucidate how a tissue can self-organize its shape.\\
Cell growth in plants largely results from yielding of cell wall under internal turgor pressure \cite{lockhart1965analysis,ray1972role}. The directional yielding of cell walls due to their anisotropic properties is behind the anisotropic growth of the plant cells. It has been long observed that cellulose microfibrils of the cell wall are oriented in transverse direction in elongating cells \cite{green1962mechanism}. The microfibrils, which are bound together by hemicelluloses and are embedded in a matrix of pectin, are the major load bearing component of the cell wall \cite{mirabet2011role}. The stiffness of the wall depends on the orientation of the fibers and is higher in the direction parallel to the orientation \cite{kerstens2001cell}. This is crucial in promoting anisotropic cellular growth from an isotropic force arising from turgor pressure. \\
Cortical microtubules (CMTs), present in the cell cortex, are decisive in the deposition of new microfibrils on the cell wall as they mediate the movement of cellulose synthase complexes \cite{ledbetter1963microtubule,heath1974unified}. The complexes move along the tracks lined by CMTs and align the cellulose microfibrils along the directions of microtubules \cite{paredez2006microtubules,somerville2006cellulose}. The orientation of CMTs itself is strongly linked with mechanical stress on the walls \cite{wymer1996plant,cleary1993pressure,hamant2008developmental,uyttewaal2012mechanical}. The microtubules generally align towards the direction of maximal stress which results in paving of cellulose microfibrils in the same direction  \cite{hamant2008developmental,uyttewaal2012mechanical,landrein2013mechanical}. As a result, stress patterns emerging during development are a putative key actor to organize growth and shapes of tissues in plants.\noindent
\begin{figure*}[hbt!]
  \includegraphics[width=\textwidth]{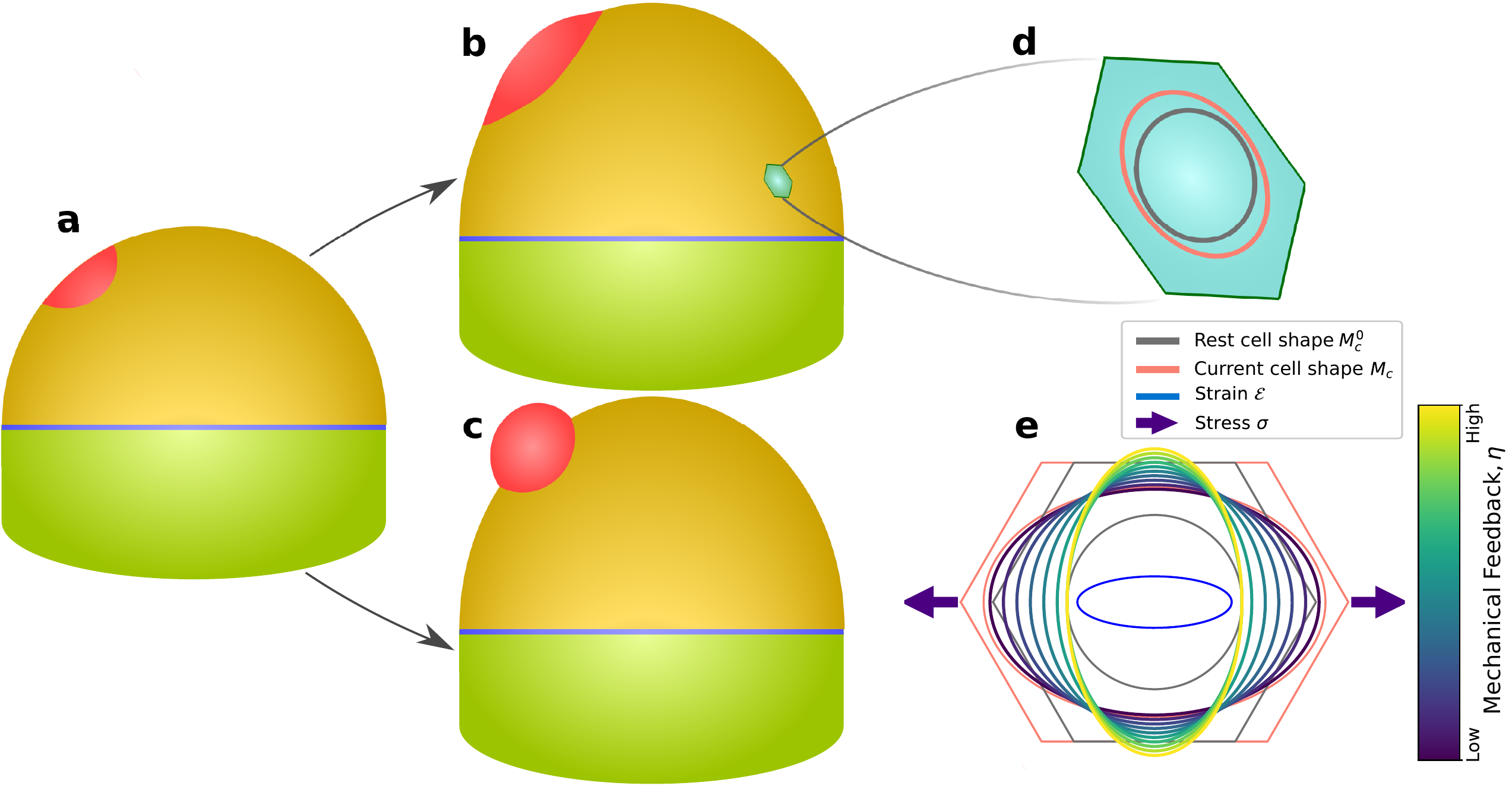}
  \caption{The outgrowth of aerial organs in plants start from primordia on the shoot apical meristem (SAM). (a)--(c) show the growth of primordia from a group of faster growing cells (red) on the SAM. (a) The SAM (yellow) is taken to be initially shaped like a dome. Red patch on the SAM shows the region of incipient primordium. The vertices at the boundary of SAM (blue line) are taken to be fixed as they are connected to hardened shoot (green base). The faster growth of primordia cells can either spread out on the surface of SAM, show in (b), or it can bulge out of the SAM, like in (c), laying the foundation for organogenesis. (d) The polygonal cells that make up the SAM dome for the simulation are defined by two key shape matrices.The rest cell shape matrix $M^0_c$ (grey ellipse) is the rest shape of the cell that it would attain without neighbouring cells or inner cells pushing outward. The current cell shape matrix $M_c$ (red ellipse) is the deformed shape of the cell observed in the tissue. (e) The anisotropic growth of the cells depends on the yielding of cell wall and mechanical stresses on the cells. The mechanical feedback inhibits the growth in higher stress direction and boosts the growth in orthogonal direction. The growth of a cell's rest shape $M^0_c$ under anisotropic stress with varying mechanical feedback is shown here. The highlighted ellipses (colors from the color bar) show how the rest cell shape will look like in the following time step for a given strength of mechanical feedback. The cell is initially hexagonal (grey polygon) with corresponding rest cell shape shown by black ellipse. The application of anisotropic stress (direction is shown by purple arrow) deforms the cell into its current cell shape (red polygon and red ellipse). The resulting strain from the stress is shown by blue ellipse. The mechanical feedback leads the growth of the cell's rest shape $M^0_c$ to be more and more orthogonal to the stress acting on the cell (ellipses, dark blue to yellow).}
  \label{fig:maincartoon}
\end{figure*}
\newcommand{\tableendline}{\\[0.5em]\hdashline}
\renewcommand{\arraystretch}{1.}
\begin{table}[h]
{
\captionsetup[table]{name=Box}
    \begin{threeparttable}
    \captionsetup{labelfont={},font={small},justification=raggedright,
              singlelinecheck=false}
    \caption{List of symbols}
    \begin{small}
    \begin{tabular}{>{\raggedright\bfseries}p{1cm}p{6cm}}
      \toprule
      $x_i, y_i$ & x and y coordinates for a vertex $i$\tableendline
      $M^0_c$ & rest cell shape \tableendline
      $M_c$ & current cell shape \tableendline
      $h$ &  primordia height \tableendline
      $A_p$ & primordial area \tableendline
      $A_T$ & tissue surface area \tableendline
      $V_T$ & volume of tissue \tableendline
      $A_c$ & area of cell \tableendline
      $\eta$ & feedback strength \tableendline
      $\kappa_f, \kappa_s$ & input growth rates for primordial and meristematic cells \tableendline
      $\kappa^*_f, \kappa^*_s$ & measured growth rates for primordial and meristematic cells  \tableendline
      $\gamma$ & fluctuations strength of growth rates \tableendline
      $r_g$ & growth ratio; $r_g = \kappa^*_f/ \kappa^*_s$ \tableendline
      $\mu$  & elastic stiffness \tableendline
      $\mu_b$ & bending stiffness \tableendline
      $\epsilon$ & strain on a cell \tableendline
      $\sigma$ & stress on a cell \tableendline
      $\sigma_{r,o}$ & stress on radial and orthoradial direction \tableendline
      $g_{r,o}$ & measured growth of cells on radial and orthoradial direction \\\bottomrule
    \end{tabular}
    \end{small}
    \end{threeparttable}
}
\end{table}\\\noindent
Besides the elastic yielding and restructuring of cell wall under stress, the patterns of cellular growth in plants are driven by biochemical signaling \cite{cosgrove2015plant}. In expanding walls, the cellulose microfibrils slide past each other under stress in a process of polymer creeping that leads to irreversible growth \cite{marga2005cell,cosgrove2005growth}. An important group of hormones that plays a major role in this growth process, by loosening up the cell walls, is auxin \cite{ray1962cell,nakayama2012mechanical,majda2018role}. Organ formation in plants at the shoot apical meristem is preceded by accumulation of auxin \cite{reinhardt2003regulation,benkova2003local,heisler2005patterns,heisler2005patterns,kwiatkowska2008flowering}. Initial outgrowth of organs from the dome shaped shoot apical meristem (SAM), called the primordia, is surrounded by localized auxin transporters that carry auxin into the incipient region \cite{reinhardt2003regulation,heisler2010alignment}. The accumulated auxin locally promotes growth in cells and initiates the formation of the primordium. During this development, disparate growth patterns emerge on the tissue. The cells at the boundary of primordia and meristem have slower and anisotropic expansion whereas the cells in the primordia and meristem grow isotropically \cite{kwiatkowska2004surface,aida2006morphogenesis}. The boundary also becomes saddle shaped as the primordia grow outwards from the SAM \cite{kwiatkowska2003growth,dumais2002analysis}. How these different growth patterns emerge from the initial accumulation of auxin is still puzzling. \\
In this work, we develop a three dimensional vertex model for the shoot apical meristem (SAM) to study plant organ outgrowth. The cellular growth of the SAM is locally increased to simulate the auxin-led local initiation in higher growth rates and the resulting primordial growth is studied. We find that the cellular ability to sense and respond to mechanical stresses within the tissue leads to efficient growth of a new primordium out of SAM. We further show that mechanical feedback on cellular growth is not only responsible for emerging pattern of growth in SAM but is also involved in redistributing the stresses acting on the cells.\\
\section*{Materials and Methods}
\subsection*{Three-dimensional Vertex Model}
Vertex models have been used to explore tissue shape in epithelial morphogenesis in a variety of model systems \cite{hamant2008developmental,boudon2015computational,farhadifar2007influence,osterfield2013three,misra2016shape,alt2017vertex,fozard2013vertex}. A vertex model represents cells as a collection of vertices
that describe their shape. They can be modelled as a polygon in two-dimensional or three-dimensional space. The cells may be in addition given a thickness by adding a height term. In
our formalism, we instead use bending stiffness of cellular layer
to represent the tissue mechanical impact of their height.\\ The vertices are shared between the neighboring
cells and this provides a vital advantage in modeling plant cells as
they share cell walls and do not slide past each other. Each of
these vertices represent a junction between cells and is subject
to force balance. The movement of vertices, representing deformation
of cells, arise from changes in this force balance due to processes like cellular
growth and cell division. The cells in our computational
model are two-dimensional polygons but are free to move around
in three-dimensional space. This allows us to investigate how individual cell growth dynamics can drive plant organ outgrowth. 
\subsection*{Shape matrices as cell representation}
To account for anisotropic cell growth, we describe the cells by a form matrix that is computed as a second moment of area matrix $M$ \cite{alim2012regulatory,uyttewaal2012mechanical}. The matrix is calculated with respect to the intrinsic coordinate system for
each cells with its components given by
\begin{equation}\label{eq:secondmomentofarea}
\begin{split}
&M_{xx} = \frac{1}{12} \sum_{i=1}^{n} a_i(y^2_i + y_iy_{i+1} + y_{i+1}^2) \\
 &\begin{multlined}
    M_{xy} = M_{yx} = \frac{1}{12} \sum_{i=1}^{n} a_i(x_iy_{i+1} +2x_iy_{i} \\
            + 2x_{i+1}y_{i+1} +x_{i+1}y_i)
\end{multlined}\\
&M_{yy} = \frac{1}{12} \sum_{i=1}^{n} a_i(x^2_i + x_ix_{i+1} + x_{i+1}^2)
\end{split}
\end{equation}
with $a_i = (x_iy_{i+1}-x_{i+1}y_i)$, where $x_i$ and $y_i$ are coordinates of a vertex $i$ measured along the intrinsic axes in the $x$ and $y$ direction, respectively.  Like an elastic line under tension, there
is a rest shape and a deformed shape for each cell. The rest shape is
the shape that a cell $c$ wants to acquire in order to reach its energy minimum and is denoted by $M^0_c$.
As cells reside in a tissue, they are pushed and pulled
from neighboring cells. The shape that a cell is deformed into, the cell's current shape $M_c$, is the one that we observe in the tissue. The energy minimum for the whole tissue lies away from the individual minimum of each cells due to the intercellular coupling. This implies that all
of the cells face some deformation of their rest shape in tissue's equilibrium
state. \\

\subsection*{Mechanical energy of tissue}
The morphology of a tissue is a result of the competition between the mechanical equilibration of the system and active biological processes inside that push it out of equilibrium. The mechanical energy for equilibration can be written as a functional with sum over costs for specific mechanical deformations. We take the functional for SAM as
\begin{equation}
	U = U_\text{elastic}+ U_\text{bending} + U_\text{pressure} \text{,}
    \label{eq:totalEnergyEquation}
\end{equation}
accounting for cell's elastic deformation, bending and the plant shoot's internal pressure as discussed in detail in the following subsections. This functional is minimized to obtain the equilibrium shape of the tissue.  
\subsubsection*{Strain Energy for the cells}
As the cells are described as polygon embedded in three-dimensions, we take a generalized relation of stresses and strains in three-dimensions using the directional information of strain tensor and the Kronecker delta tensor, $\delta_{ij}$,
\begin{equation}
	\sigma_{ij} = 2\mu \epsilon_{ij} + \lambda \delta_{ij} \sum_k\epsilon_{kk}\text{.}
    \label{eq:lameParameterization}
\end{equation}
The parameters $\lambda$ and $\mu$  are Lamé's first and second parameter, respectively. An elastic energy density for any deformation of an isotropic material is thus written as 
\begin{equation}
    \upsilon_\text{elastic} = \mu\sum_{ij} \epsilon^2_{ij} + \frac{1}{2}\lambda \Big(\sum_i \epsilon_{ii}\Big)^2 \text{,}
    \label{eq:elasticEnergySurfaceDensity}
\end{equation}
using Eq.~\ref{eq:lameParameterization}. To find the strain energy expression for the vertex model, strain and stress tensors need to be defined in terms of the shape matrices that are used to describe the cells (Eq.~\ref{eq:secondmomentofarea}). Strain can be expressed as the difference between current shape and initial shape, written as
\begin{equation}
    \epsilon_c = \frac{M_c-M^0_c}{\Tr(M^0_c)} \text{.}
    \label{eq:strainVertexModel}
\end{equation}
The stress can then be defined using Eq.~\ref{eq:lameParameterization}. With these definitions, a complete expression for the elastic energy can be calculated by integrating Eq~\ref{eq:elasticEnergySurfaceDensity} over the tissue surface to obtain 

\begin{equation}
	U_\text{elastic} = \mu \sum_c A_c \frac{\Vert{M_c-M_c^0\Vert}^2_2}{tr^2(M^0_c)}
	+\frac{1}{2}\lambda \sum_c A_c\frac{tr^2(M_c-M_c^0)}{tr^2(M^0_c)} \text{,}
    \label{eq:elasticEnergyVertexModel}
\end{equation}
where $A_c$ is the area of a cell $c$.
We set $\lambda = 0$, which is proportional to the Poisson ratio, to further simplify the elastic energy expression to 
\begin{equation}
	U_{elastic} = \mu \sum_c A_c\frac{\Vert{M_c-M_c^0\Vert}^2_2}{tr^2(M^0_c)} \text{.}
    \label{eq:elasticEnergyVertexModelSimplified}
\end{equation}
This simplification has no impact on the simulation results as the mechanical behavior in developing tissue can be considered stable under varying Poisson ratio \cite{bozorg2014stress}. Our simulations prove to be qualitatively robust against variations in Poisson ratio, see Supplemental Material S14.\\
The stress can then be explicitly expressed in terms of the cell shapes as 
\begin{equation}
    \sigma_c = 2\mu\frac{M_c-M^0_c}{\Tr(M^0_c)}\text{.}
\end{equation}

\subsubsection*{Bending energy of the tissue}
Previous authors have noted the response of the shoot apical meristem is close to a stiff shell inflated by a pressure \cite{beauzamy2015mechanically}. This suggests that the turgor pressure from within the tissue is sustained by either the outer layer of cells or only the outer walls of those cells.  We, thus, consider the meristem as a single layer of stiff cells on a two-dimensional surface, free to move in three-dimensional space. For epithelial cells in a tissue, the cells are restricted by, first, the walls that are perpendicular to the surface (anticlinal walls) and, second, by junctions with cells around them. Any significant bend or twist away from the epithelial surface would mean a major deformation on the anticlinal walls and on cells underneath. Thus, we add a bending term to the mechanical energy that penalizes deformations of anticlinal walls. It is based on works of Canham and Helfrich, who considered a three-dimensional soft object with an infinitely thin interface with bending resistance \cite{canham1970minimum,helfrich1973elastic,guckenberger2016bending},
\begin{align}
	U_{bending} = 2 \mu_b \int_S &(H-H_0)^2 dA+ \int_S \mu_K K dA \text{,}
    \label{eq:helfrichBendingComplete}\\
    H &= \frac{1}{2} (k_1 + k2)\text{,}
    	\label{eq:meanCurvature}\\
    K &= k_1k_2\text{,} \label{eq:gaussianCurvature}
\end{align}
where, $H$ is the local mean curvature and $K$ is the Gaussian curvature. $k_1$ and $k_2$ are the principal curvatures at a point on the tissue surface $S$. $H$ is taken to be  positive for the dome shape of the shoot tip. The Gaussian curvature $K$ can be integrated out of the energy equation as it remains constant for a surface with fixed topology, which leaves only the first term for bending energy \cite{gompper1996random,meyer2003visualization}. Discretization of $H$ developed by Meyers et al.~is used to compute Eq.~\ref{eq:helfrichBendingComplete} for the tissue in our simulations \cite{meyer2003visualization,guckenberger2016bending}.
\subsubsection*{Pressure inside the tissue}
The cells below the surface epithelial layer of the shoot apex push outwards on the surface layer. The net force acting on the cells in the surface layer promotes outward growth. Following previous approaches \cite{hamant2008developmental,louveaux2016cell,boudon2015computational} we represent this outward pressure by an additive pressure term in the energy,
\begin{equation}
	U_{pressure} = - P V_T \text{,}
    \label{eq:pressureEnergy}
\end{equation}
where, $P$ is the pressure from underneath and $V_T$ is the volume of the total shoot apex. Note, that the contribution of internal pressure within individual cells can be subdivided into a perpendicular and and in-plane contribution. Under the assumption of equal pressure in all cells the in-plane contribution cancels out \cite{hamant2008developmental}, the perpendicular component has the same functional form as above and as such reflected in the term as well.
\subsection*{Energy minimization and boundary condition}
The equilibrium shape of the tissue is found by minimization of the mechanical energy Eq.~\ref{eq:totalEnergyEquation} using the SubPlex algorithm implemented in the open-source non-linear optimization library NLOPT \cite{rowan1990subplex,nloptJohnson}. During the entire simulation the vertices at the lower boundary of the tissue are fixed in their position (Fig.~\ref{fig:maincartoon} $a-c$) representing the connection of SAM to mature and hardened cells of the shoot. 
\subsection*{Cellular growth through deformation}
The cellular growth in plants has long been regarded as a mechanical process with yielding of cell wall leading growth under turgor pressure. Lockhart considered cell wall as Bingham fluid and proposed to model a cell's growth proportional to the deformation on the cells \cite{lockhart1965analysis}. Adapting this definition for the vertex model we write
\begin{equation}
	\frac{dM^0_c}{dt} = \kappa(1+\gamma) (M_c-M^0_c)_+ \text{,}
\label{eq:restShapeLockhartGrowth}
\end{equation}
where, $\kappa$ is the growth rate of cells, fluctuating with amplitude $\gamma$. $\gamma$ represents the variation arising in the cellular growth including the variations in turgor pressure present between the cells that might result in inhomogeneous growth. The difference of the current shape, $M_c$, and the rest shape, $M^0_c$, i.e.~the deformation on the cells drives the growth. The operation $(.)_+$ ensures that the cells do not shrink even if the cells are faced with negative deformation under compressive stresses (Eq.~\ref{eq:positiveSemiDefiniteCriteria}). This operation only allows positive growths led by positive deformations.
\subsection*{CMT-led mechanical feedback on cell wall}
\label{section:feedback}
\begin{figure}[hbt!]
\centering
	\includegraphics[width=0.6\linewidth]{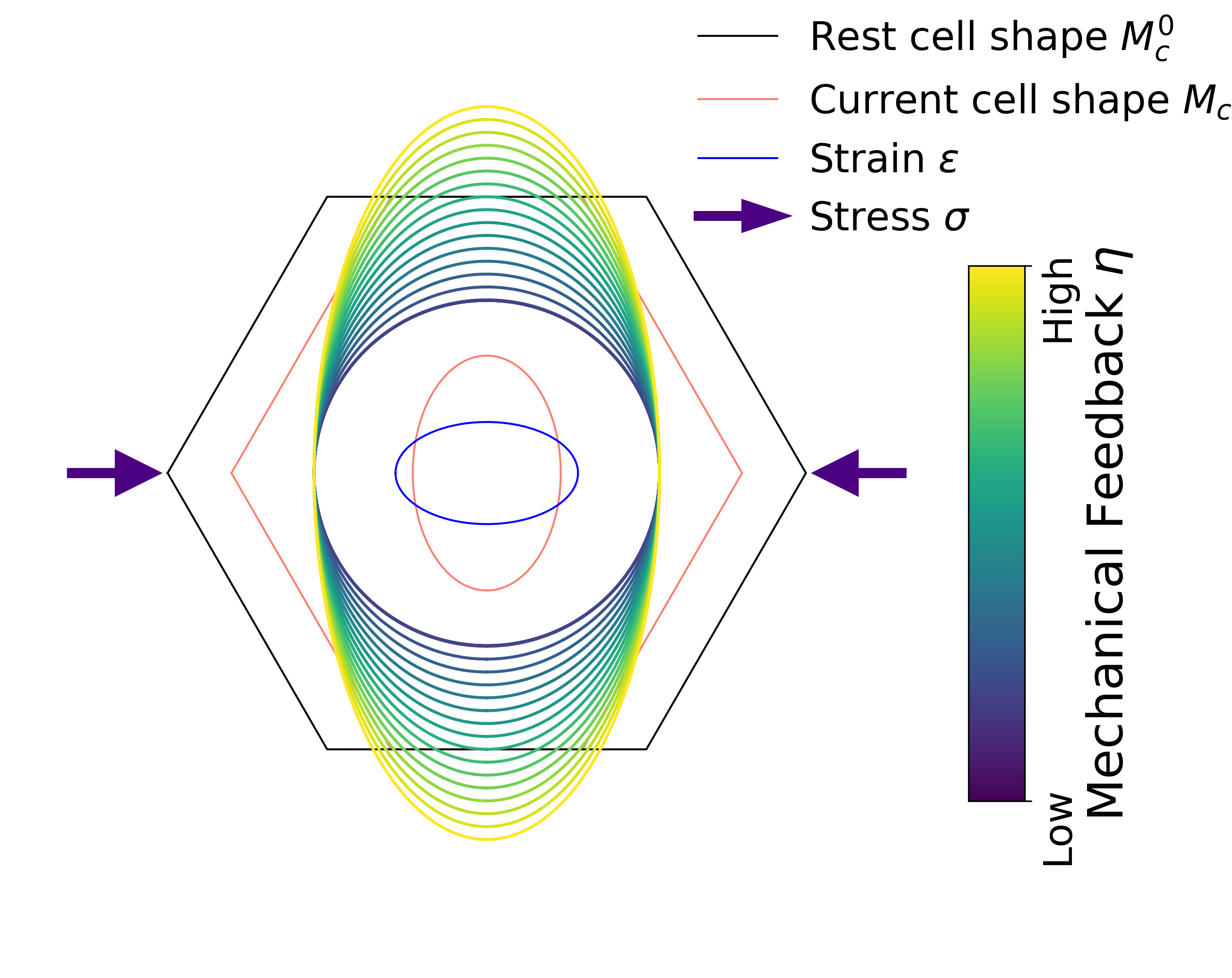}
    \caption{The response of the cellular growth form the mechanical feedback is identical for compressive stress as compared to the tensile stress as shown in Fig.~\ref{fig:maincartoon} $e$. The rest cell shape (grey ellipse) is compressed to the current cell shape shown by red ellipse due to the stress denoted by purple arrow. As a response, the microtubules are expected to align in the higher stress direction (horizontal) strengthening the walls. This results in growth that is increasingly orthogonal to stress direction as shown by the ellipses from dark blue to yellow with increasing mechanical feedback.}
    \label{fig:compressiveFeedback}
\end{figure}

The anisotropic cellular expansion and growth patterning in plants depend on the anisotropic cell wall stiffness as the forces generating growth are isotropic. The reorganization of CMT orientation, led by stresses, and the subsequent cellulose microfibril deposition promoting wall anisotropy can be represented by the dynamics of the rest cell shape \cite{alim2012regulatory,uyttewaal2012mechanical}.
Given the observation that CMT orient according to the highest stress and thus reduce growth in the direction of highest stress, we model this effect by coupling the growth rate to the cell's asymmetric stress component, the deviatoric stress $D_c = \sigma_c - 1/2\Tr{\sigma_c}$, thus extending the growth equation Eq.~\ref{eq:restShapeLockhartGrowth} to\\
\begin{equation}
\begin{split}
	\frac{dM^0_c}{dt} = & \kappa(1+\gamma) (M_c-M^0_c) \\
    &-\frac{\eta}{2} \Big(D_c(M_c-M^0_c) +
    (M_c-M^0_c)D_c\Big) \text{.}
\end{split}
\label{eq:restShapeLockhartGrowthFeedback}
\end{equation}
The feedback parameter $\eta$ quantifies the rate of rest shape reorganization per unit of stress for a cell. It represents the cell wall's ability to respond to stress and with higher $\eta$, the efficiency of reorganizing of  the cell walls is higher. Increasing mechanical feedback results in growth that is more and more orthogonal to the higher stress direction, as expected from the wall strengthening in that direction (Fig.~\ref{fig:maincartoon} $e$ and Fig.~\ref{fig:compressiveFeedback} in the Supporting Materials). This anisotropic growth of the cells by the modulation of rest cell shape with mechanical feedback takes into account the anisotropic properties of cell walls which is not directly included in the elastic energy density.\\
The reorganization of the wall stiffness in the cells from the mechanical feedback can be measured by comparing the growth of the cells to its deformation. Supplemental Material~\ref{section:stiffnessmodulation} details the measure for the stiffness modulation of the cells.
\subsection*{Localized enhanced growth rate}
\begin{figure}[hbt!]
	\centering 
	\includegraphics[width=0.4\linewidth]{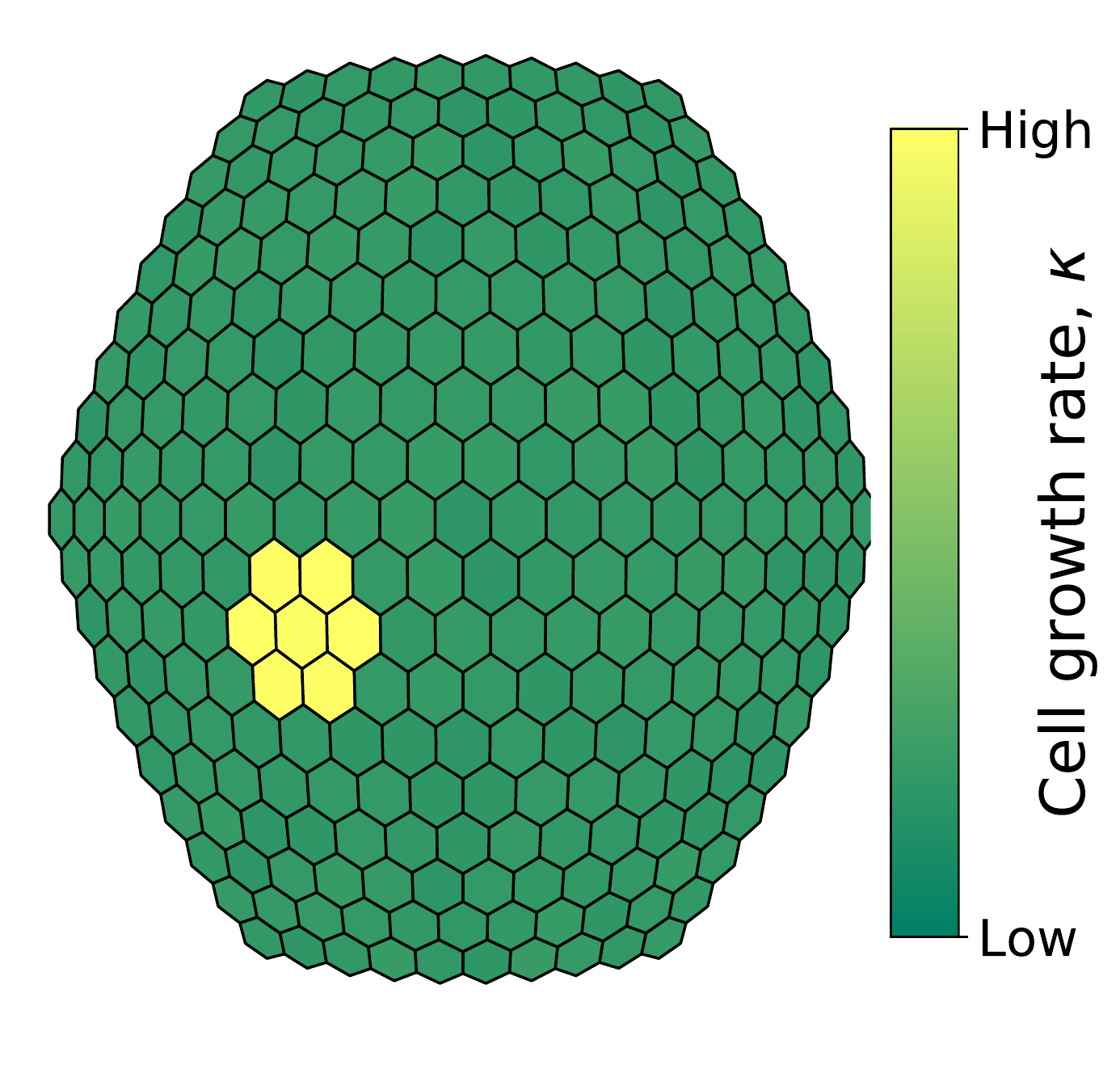}
	\caption{The localized accumulation of auxin causes an increased growth rate in primordium cells (yellow), top-down view on shoot tip. This is modelled by assigning higher growth rate to cells of designated primordial region.}
	\label{fig:auxinAccumulation}
\end{figure}

The plant growth hormone auxin causes reduction in cell wall hemicellulose polysaccharides, increase in pectin polymerization and viscosity, among other roles in the plant biology \cite{ray1962cell,nishitani1981auxin,majda2018role,perrot2010cellular}. It initiates organ formation on the SAM by increasing the growth rate of primordial cells through loosening of the cell walls \cite{reinhardt2003regulation,nakayama2012mechanical,hamant2008developmental}. Yet, the faster growing cells in primordial region are still tightly connected to the slower growing cells in the meristem tissue through shared cell walls \cite{rebocho2017generation,hamant2008developmental,boudon2015computational}. Thus, it is unclear how fast both primordial and meristem cells can effectively grow and how both kinds of cells deform due to the localized enhanced growth rate. To study tissue growth and deformation, we define a prepatterned localization of auxin in the SAM (Fig.~\ref{fig:auxinAccumulation}) with an enhanced growth rate $\kappa_f$ relative to the surrounding meristem tissue with $\kappa_s$ as input parameters to Eq.~\ref{eq:restShapeLockhartGrowthFeedback}. Due to the cell-cell junctions and tissue mechanics constraining the cells, the actual growth rates of cells, $\kappa^*$, is less than specified by the input parameters. The rates $\kappa^*_{f}$ and $\kappa^*_{s}$ are measured in simulations by fitting an exponential growth curve to the area growth of primordial and meristematic cells, respectively (Fig.~\ref{fig:varyfkexponentialfit}).  Ultimately, the ratio
\begin{equation}
    r_g = \kappa^*_{f}/\kappa^*_{s}
    \label{eq:growthRatio}
\end{equation}
of these two growth rates is what is governing the growth rate of the entire tissue. The two values ($\kappa^*$ and $\kappa$) are not equal as the cellular growth is effected by tissue mechanics, cellular interactions and mechanical feedback.
\\\noindent
The chosen parameters for the simulation are listed in Supplemental Material Table~\ref{tab:simulationparameters}. The system is robust under parameter change up to two orders of magnitude for pressure and stiffness, while the growth ratio larger than $10$ can lead to strong artifacts in cell shapes.
\section*{Results}
\subsection*{Growth of shoot apical meristem }
\begin{figure}[hbt!]
\centering
        \includegraphics[width=0.8\linewidth]{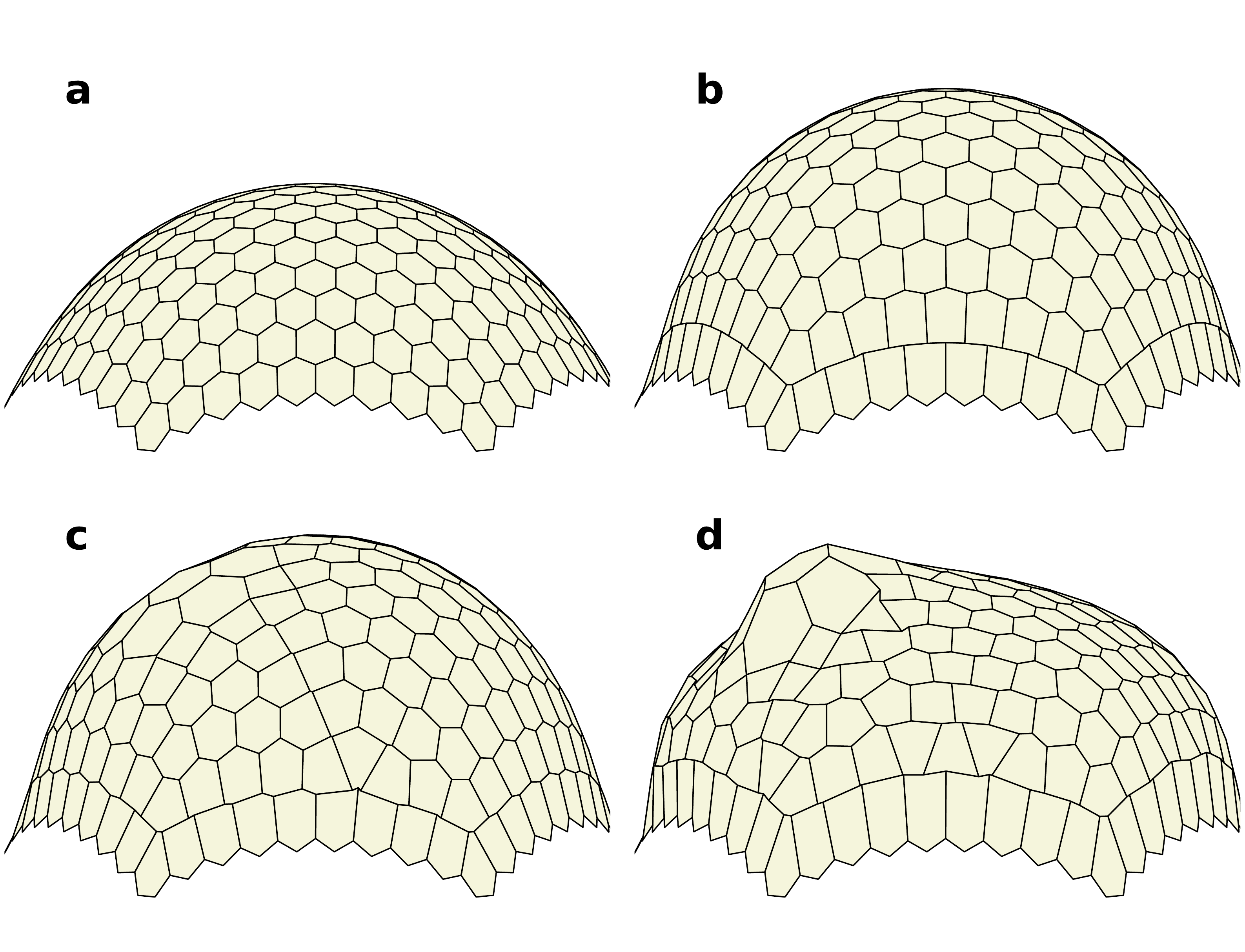}
	\caption{Growth of shoot apical meristem under varying growth conditions. (a) The initial shape of a tissue used for all growth simulation with surface area  $A_T = 665$.  (b), (c) and (d) are the resulting shape after tissue has grown to $A_T = 850$. (b) Growth with uniform growth rate $\kappa = 0.5$ (c) Growth with growth ratio $r_g = 4.8$ and no mechanical feedback $\eta = 0$. (d) Growth, also with growth ratio $r_g = 4.8$ but high mechanical feedback $\eta = 8$. }
	\label{fig:lockhartHomogeneousGrowth}
\end{figure}

\begin{figure}[hbt!]
\centering
	\includegraphics[width=0.8\linewidth]{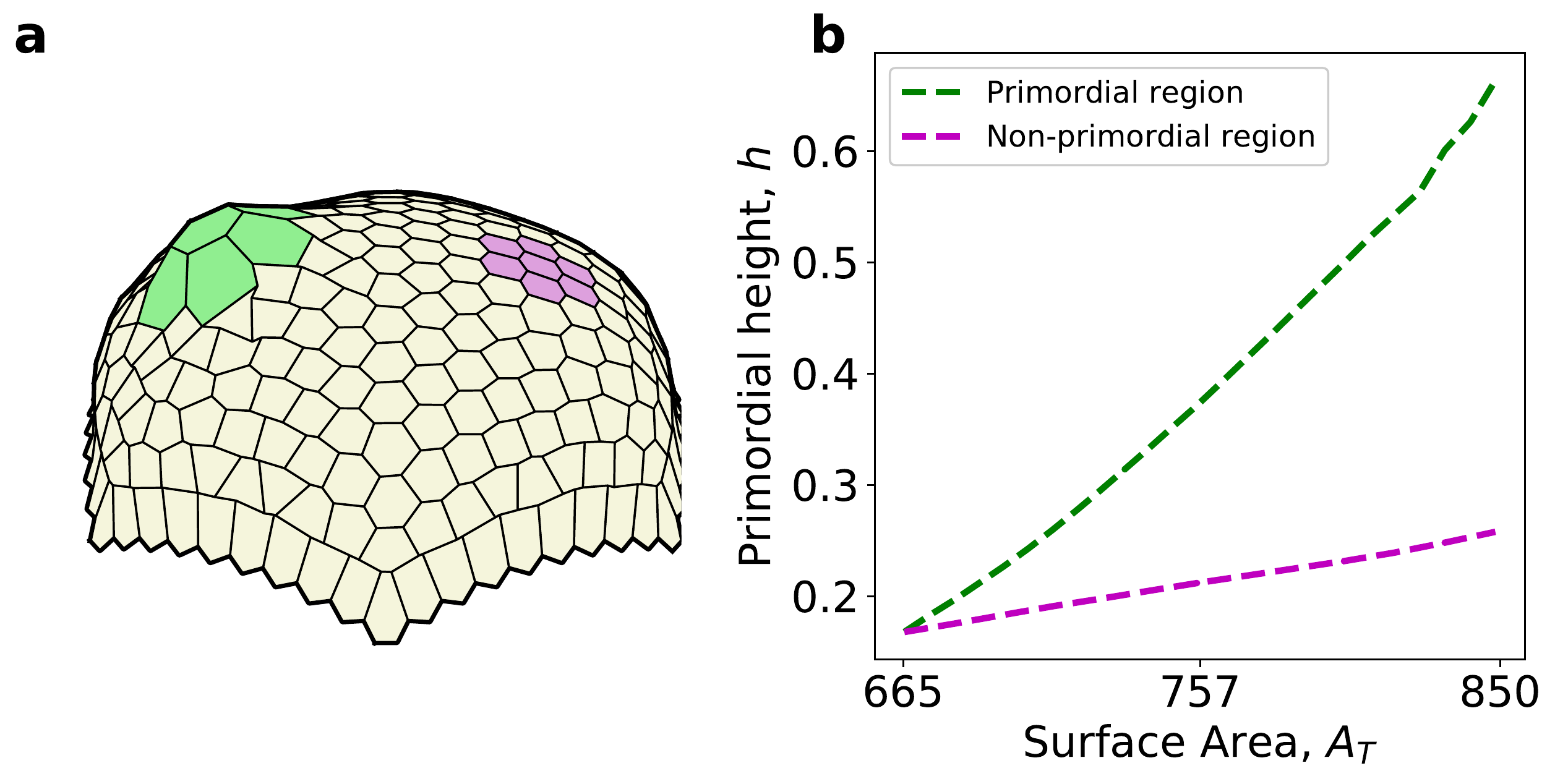}
	\caption{Comparison of tissue height in primordial and non-primordial region shows the presences of significant outgrowth in primordial region. (a) Regions of primordia (green) and non-primordia (magenta) shown on the tissue. (b) Increase in height of primordial and non-primordial region.}
	\label{fig:primiordia-nonprimiordia-ChangeComparision}
\end{figure}

Aerial organs in plants start out as primordia in the SAM, initiated by differential growth of cells. During the emergence of primordia, cells in primordia are observed to grow faster and isotropically whereas the cells in boundary region between the primordia and the rest of the meristem have arrested growth and are highly anisotropic \cite{reddy2004real,kwiatkowska2003growth,aida2006morphogenesis}.
 To understand the cause of these growth pattern, the overall role of mechanics-led growth feedback and their effect on primordium outgrowth, we developed a three-dimensional vertex model to simulate the growth of SAM.\\
We take the SAM as a hemispherical surface composed of homogeneous hexagonal cells that have been relaxed under the chosen simulation parameters. With an uniform cellular growth rate $\kappa$ (see Eq.~\ref{eq:restShapeLockhartGrowth}) for all cells, the tissue expands without significant morphological changes on the surface  (Fig.~\ref{fig:lockhartHomogeneousGrowth} $a$ and $b$). The growth of the SAM is driven by the deformation of the surface cells due to the volume pressure from the tissue underneath.\\
Plant organ outgrowth on the SAM is observed when the tissue is prepatterned with a localized higher growth rate corresponding to localized auxin accumulation in primordial cells (Fig.~\ref{fig:auxinAccumulation}). The faster growing region bulges out from the tissue surface as shown in Fig.~\ref{fig:primiordia-nonprimiordia-ChangeComparision} and Fig.~\ref{fig:lockhartHomogeneousGrowth} $c-d$. We quantify this outgrowth of the primordium by measuring the height of the bulge as
\begin{equation}
	h = \Vert{\vec{v}_{\text{top}} - \vec{v}_{\text{boundary}}}\Vert \text{,}
	\label{eq:heightOfPrimordium}
\end{equation}
where, $\vec{v}_{\text{boundary}}$ is the average position of the vertices at the boundary of the primordial region and $\vec{v}_{\text{top}}$ is the position vector to the centroid of the cell at the top of the primordium, as shown in Fig.~\ref{fig:growthHeightMeasurement}.\\
We analyze the outgrowth height as a function of tissue surface area to facilitate a comparison independent of the chosen intrinsic cell growth rates. The total simulation time and cellular growth over one time step can differ significantly depending on the choice of growth rates. However, as all of the dynamics in biology follow a robust timescale for the growth, we use the growth of the tissue surface area as an indicator for time to compare simulated tissues under different sets of parameters.\\ 
In the following sections, we examine the simulation results of organ outgrowth to investigate the role of tissue mechanics during primordial growth.

\subsection*{Differential growth leads to primordial outgrowth}
\begin{figure}[hbt!]
\centering
		\centering
		\includegraphics[width=0.8\linewidth]{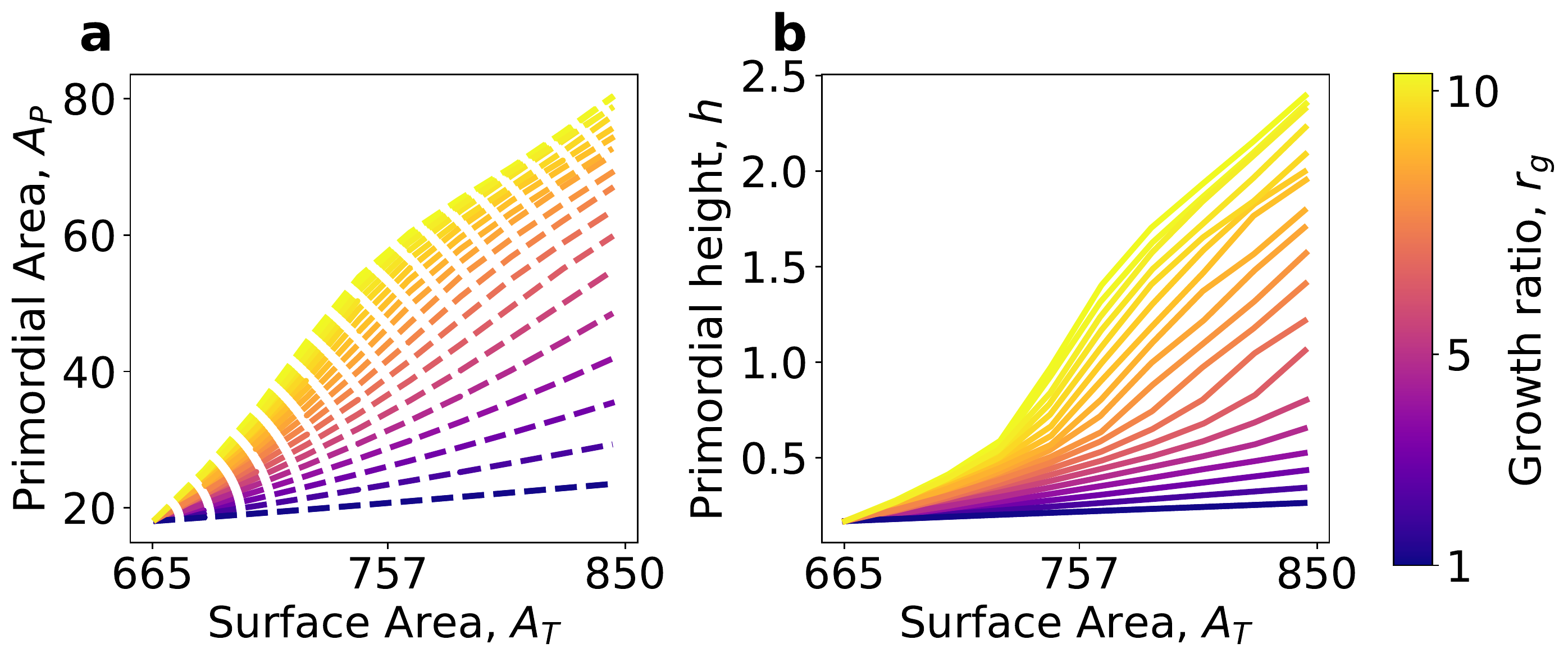}
	\caption{Higher growth ratio leads to higher primordial growth. The growth rate for meristem ($\kappa_{s}$) is kept constant while the primordial growth rate ($\kappa_f$) is increased for larger growth ratio. (a) The primordium grows larger with increasing growth ratio. (b) The primordium bulges out further due to the increase in its size as seen with higher primordial height on greater growth ratio. }
	\label{fig:growthRatioChange}
\end{figure}

The faster growth rates of primordial cells pushes against the SAM surface leading to the bulging out of the tissue. Here, the ratio of growth rates $r_g$ (Eq.~\ref{eq:growthRatio}) dictates bulge formation. Increasing $r_g$ leads to higher outgrowth, see Fig.~\ref{fig:growthRatioChange} $b$, going hand in hand with stronger growth of primordial cells and the formation of bigger sized bulge (Fig.~\ref{fig:growthRatioChange} $a$). Changing both the primordial and meristematic growth rates while keeping the growth ratio constant has no effect on the height dynamics (Fig.~\ref{fig:growthRationChangeComparision}). To further explore the emergence of primordia, we next introduce  a mechanical feedback on cellular growth to tissue wide mechanical stresses and study its impact on outgrowth dynamics. \\
\subsection*{Active mechanical response from cells drives outgrowth}
\begin{figure}[hbt]
	\centering
	\includegraphics[width=0.8\linewidth]{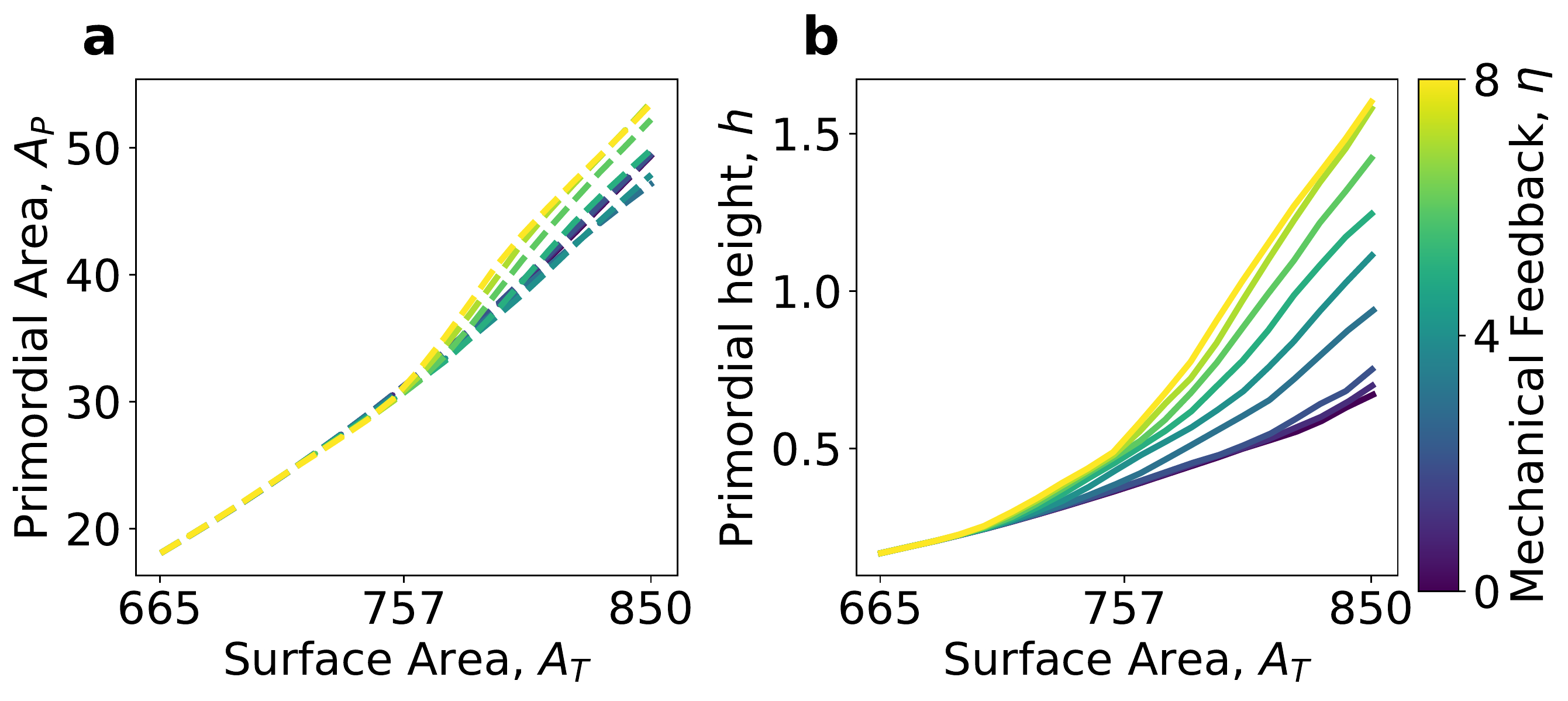}
	\caption{Increasing mechanical feedback of cells to tissue wide mechanical stresses results in efficient primordial growth. Here, a tissue with growth ratio $r_g = 4.8$ is grown for varying mechanical feedback. (a) The overall areal growth of the primordium is relatively unchanged with changing mechanical feedback. (b) The height of primordium increases significantly with higher mechanical feedback.}
	\label{fig:feedbackGrowth}
\end{figure}

Mechanical stresses in tissue are propagated among cells through shared cell walls. As a response to the mechanical stresses acting on them, the cells actively remodel cell walls. This microtubules-led reorganization of walls and the cellular growth is considered to be vital for robust morphogenesis.
We model this feedback by implementing a stress dependent term in the growth equation that accounts for active strengthening of walls in higher stress direction (Eq~\ref{eq:restShapeLockhartGrowthFeedback}).\\
We find that the ability of cells to sense stresses and react accordingly is vital for organ outgrowth on the meristem. By modulating the mechanical feedback of a tissue we observe that the outgrowth is higher when cellular response to mechanics is enhanced (Fig.~\ref{fig:feedbackGrowth} $b$). This observation in our simulation is in agreement with previous experimental observations \cite{hamant2008developmental,oliveri2018regulation}. \\
Note, that contrary to the dynamics for an increasing growth ratio, increasing mechanical feedback only promotes outgrowth height while leaving the primordial tissue area almost unchanged (Fig.~\ref{fig:growthRatioChange} $a$ and Fig.~\ref{fig:feedbackGrowth} $a$ ). This indicates that mechanical feedback promotes organ outgrowth by a different mechanism than effective increase in growth rate. Notably, growth rates in cells of primordial and meristematic regions are unaffected by mechanical feedback, keeping the growth ratio fixed. Thus, it is the more puzzling that the reorganization of growth led by mechanical feedback is able to bulge out the primordium more efficiently with increasing feedback. A little bit of insight can already be gained from the simulation snap shots in Fig.~\ref{fig:lockhartHomogeneousGrowth} $c-d$, where tissue of the same overall area with and without feedback are compared. The growth is directed outwards for the primordium with mechanical feedback, leading to a clear bulging (Fig.~\ref{fig:lockhartHomogeneousGrowth} $d$), while the primordial cells without feedback grow predominantly  within the meristem surface but are not able to bulge outwards (Fig.~\ref{fig:lockhartHomogeneousGrowth} $c$).
\subsection*{Diverging stresses reorganize growth in boundary cells}
\begin{figure}[hbt!]
	\centering
	\includegraphics[width=0.8\linewidth]{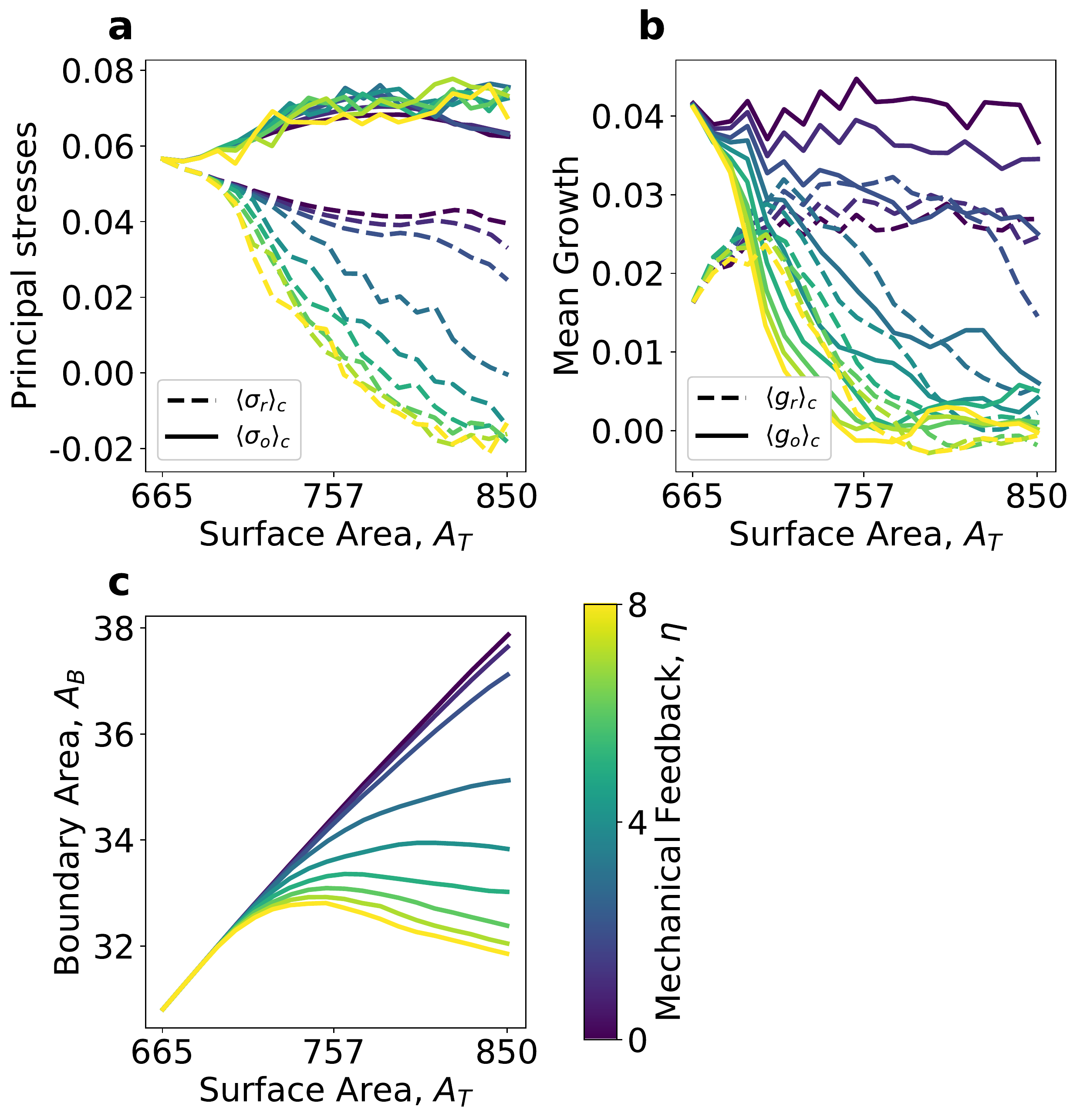}
	\caption{Pattern of stresses and growth in boundary cells undergo significant modification by mechanical feedback. Here, $\langle \cdot \rangle_c$ represents an average over the cells in the boundary of primordium.  (a) Stresses in radial ($\sigma_r$) and orthoradial ($\sigma_o$) directions diverge during growth more and more with increasing feedback. (b) Growth rates of boundary cells decay with feedback. (c) The boundary cells not only cease in growth but are also compressed by primordium and meristem cells.}
	\label{fig:feedbackROStress}
\end{figure}

\begin{figure}[hbt!]
\centering
		\includegraphics[width=0.3\linewidth]{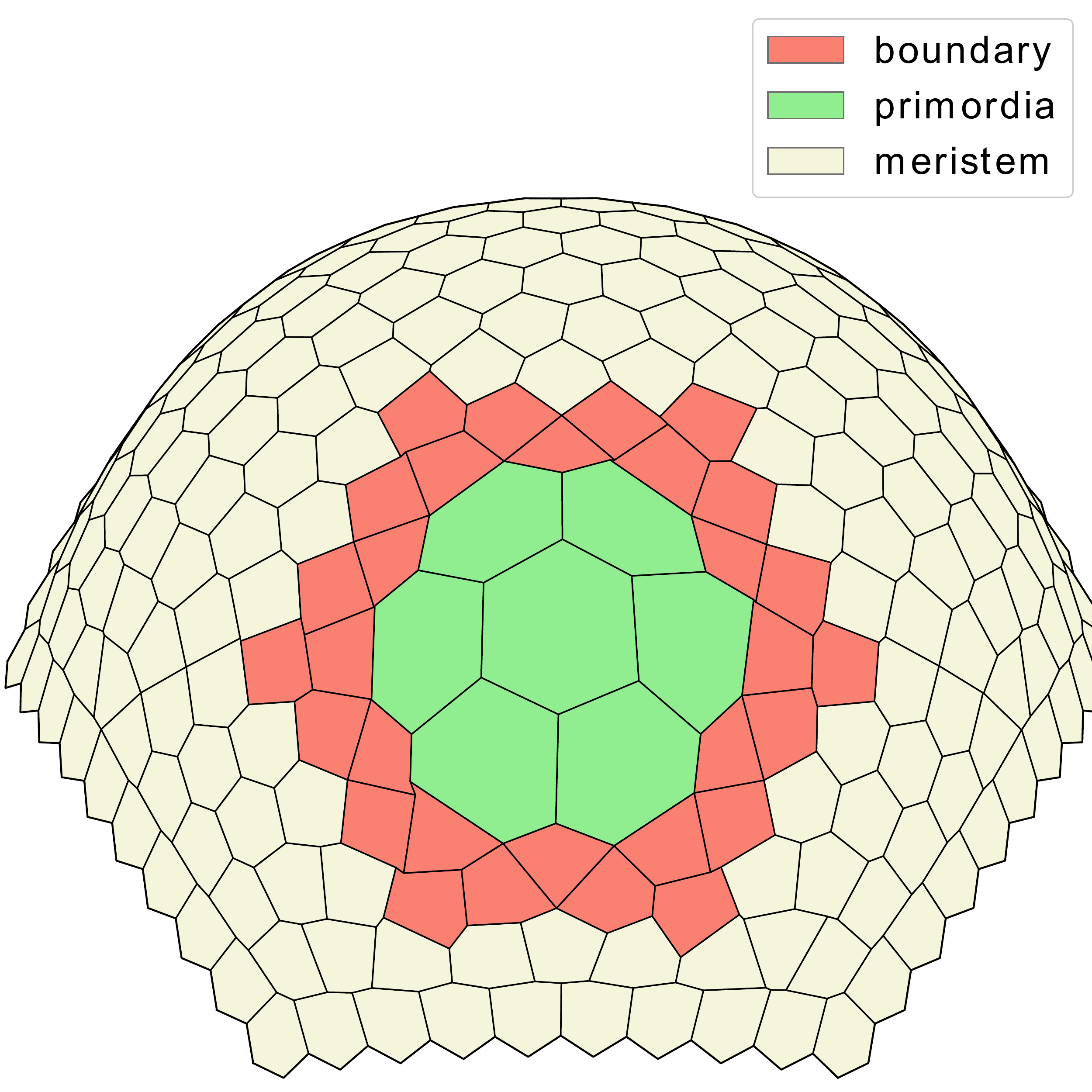}
	\caption{The meristematic, boundary and primordial cells labelled on the shoot tissue.}
	\label{fig:boundarycellslabelled}
\end{figure}

We next investigated how growth is reorganized within the tissue by the mechanical feedback and how it can lead to greater height growth in primordia. The differential growth rates between primordium and meristem reshape stress patterns on the SAM, which are used by cells, through mechanical feedback, to reorganize their growth. Mapping out the radial and orthoradial stresses on the cells at the boundary of the primordium (Fig.~\ref{fig:boundarycellslabelled}), we find that the stress distribution in boundary cells becomes more and more anisotropic during growth with increasing mechanical feedback (Fig.~\ref{fig:feedbackROStress} $a$). The orthoradial stress $\sigma_o$ (circumferential direction to the primordium) in boundary cells remains high throughout the primordial growth where as the radial stress $\sigma_r$ (direction towards the tip of primordium) declines. \\
\noindent
As expected from the mechanical feedback, the growth of the boundary cells also exhibits distinct anisotropic patterns (Fig.~\ref{fig:feedbackROStress} $b$). In the absence of feedback both orthoradial and radial growth stay at a more or less constant high level with orthoradial growth being about twice as large as radial growth. Mechanical feedback drives both orthoradial and radial growth to plummet over time to smaller and smaller growth eventually ceasing growth entirely at high mechanical feedback. The cessation of growth of the boundary cells is clearly visible when plotting the total area of boundary cells during growth, see Fig.~\ref{fig:feedbackROStress} $c$. Further, the relative stiffness reorganisation also shows the trend of growing anisotropic wall properties arising from the mechanical feedback (Fig.~\ref{fig:stiffnessReorganisationn0} and Fig.~\ref{fig:stiffnessReorganisationn8}). The stiffness is enhanced in the orthoradial direction and is lowered in the radial direction. Importantly, this enhancement is observed to be significantly boosted with the mechanical feedback. These trends of growth and mechanical patterning remained intact with the introduction of Poisson ratio on the system (Fig.~\ref{fig:varyPoissonRatio}).\\
Increasing mechanical feedback not only leads to the slower growth and stiffening of the boundary region but also to its compression (Fig.~\ref{fig:feedbackROStress} $c$) due to the increasing stresses from neighboring cells (note negative stresses arising in Fig.~\ref{fig:feedbackROStress} $a$). In addition, the shape of the boundary is also seen to be dependent on the mechanical regulation of cell growth. The Gaussian curvature, indicating the saddle shape of the boundary, is observed to be increasingly negative with feedback (Fig.~\ref{fig:gaussiancurvature} $c$).\\
We infer from these observations that the reorganization of growth and stiffness in primordium boundary cells due to mechanical feedback to the arising stresses is the vital mechanism behind the efficient outgrowth of organ primordium.\\
\subsection*{Mechanical feedback modulates the height growth rate}
\begin{figure}[hbt!]
	\centering
	\includegraphics[width=0.5\linewidth]{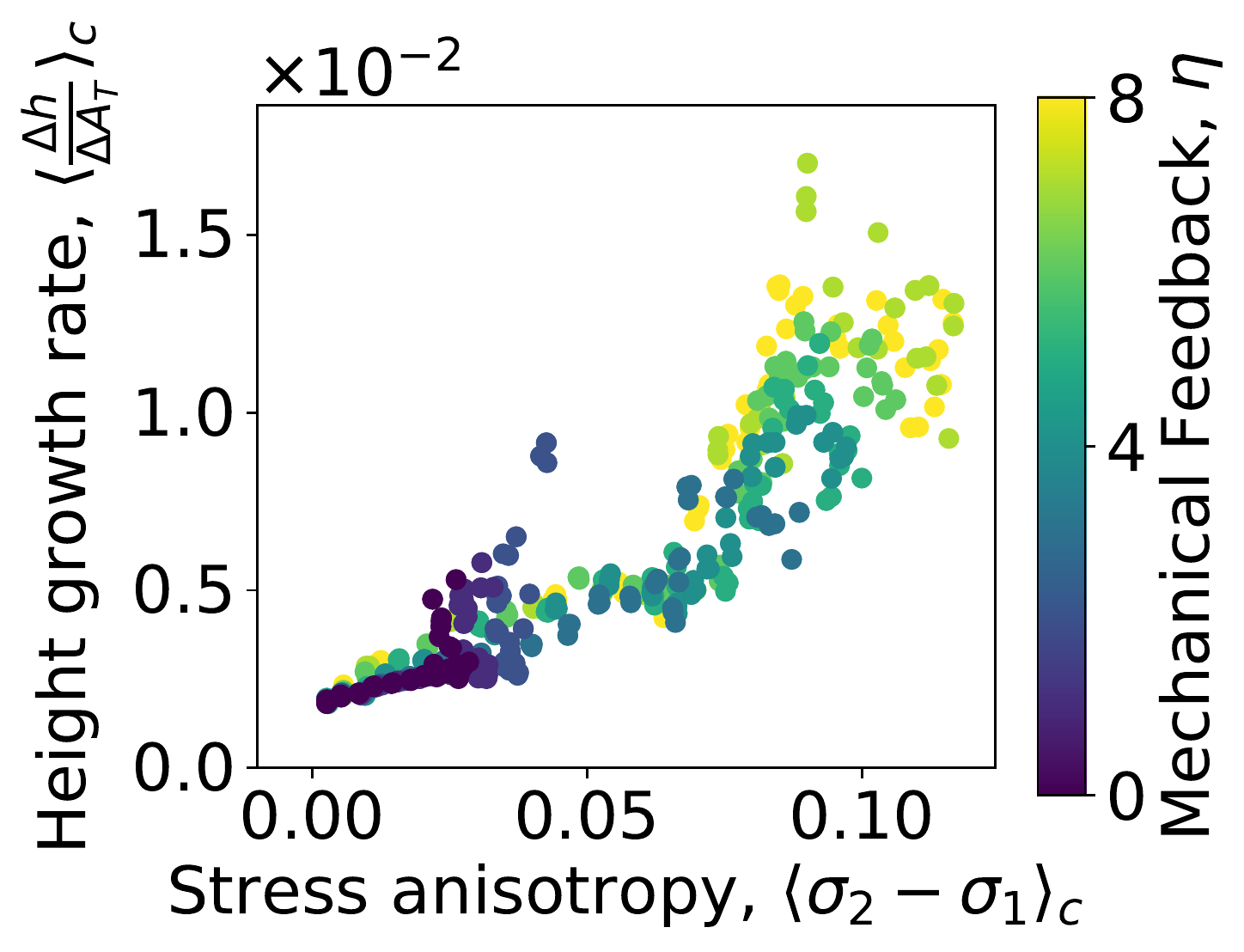}
	\caption{Rate of primordial height growth is boosted significantly by mechanical feedback. With higher mechanical feedback, both the rate of height growth and the stress anisotropy of  cells on the primordial boundary increase. By this two-way reinforcement, mechanics is able to guide efficient primordium outgrowth on the SAM.}
	\label{fig:anisotropyScatterPlot}
\end{figure}\noindent
To link the relation of generated stress pattern and growth reorganization caused by mechanical feedback, we examine the rate of height growth in the primordium with respect to the growth of the tissue surface as a function of anisotropy in stresses on boundary cells (Fig.~\ref{fig:anisotropyScatterPlot}). The stress anisotropy was defined as the difference between the two principal stresses acting on the cells. We observe that along with the boost in the height growth rate, which results directly from cellular growth reorganization, mechanical feedback generates greater stress anisotropy in boundary cells. The averaged differences in principal stresses increase with higher mechanical feedback. This helps to amplify the growth heterogeneity in the tissue and establishes the large scale stress pattern that promotes efficient organ outgrowth. Without the mechanical feedback, both height growth rate and stress anisotropy are low, see Fig.~\ref{fig:anisotropyScatterPlot}. Only the stress anisotropy in boundary cells arising from mechanical feedback, also seen by diverging stress in Fig.~\ref{fig:feedbackROStress} $a$, allows the growth reorganization in cells that results in a strong growth in primordium height.
\\
\section*{Discussion}
\begin{figure}[hbt!]
\centering
	\centering
	\includegraphics[width=0.8\linewidth]{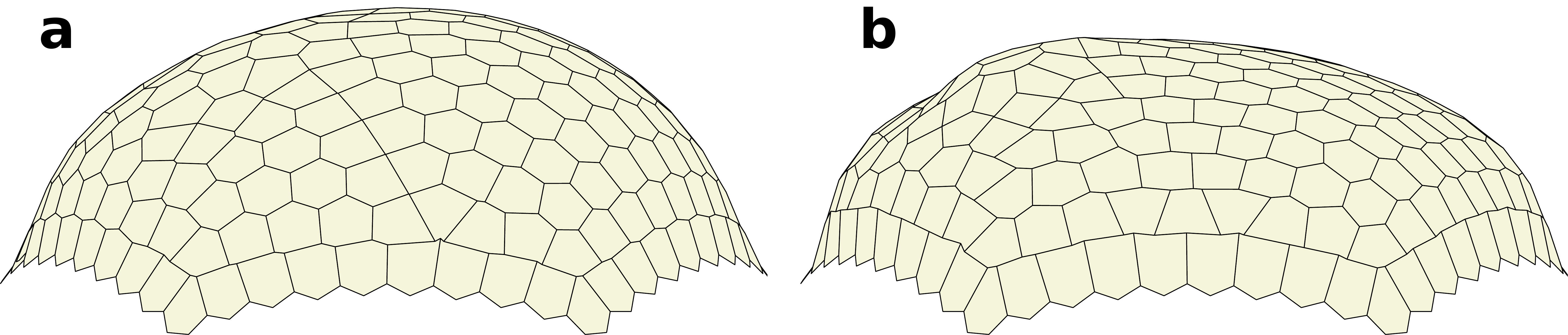}
	\caption{The effectiveness of primordial growth with mechanical feedback remains intact under the introduction of cell division. The tissue shape at surface area $A_T = 850$ for (a) $\eta = 0$ and (b) $\eta = 8$ is shown. The cells are allowed to divide here once they reach their target size (twice the initial size). New walls follow the direction of shortest axis that go through the center of the cells. Clear primordial development is seen with mechanical feedback in (b), compared to the case without mechanical feedback in (a).}
	\label{fig:cellDivision}
\end{figure}

\begin{figure}[hbt!]
\centering
	\includegraphics[width=0.4\linewidth]{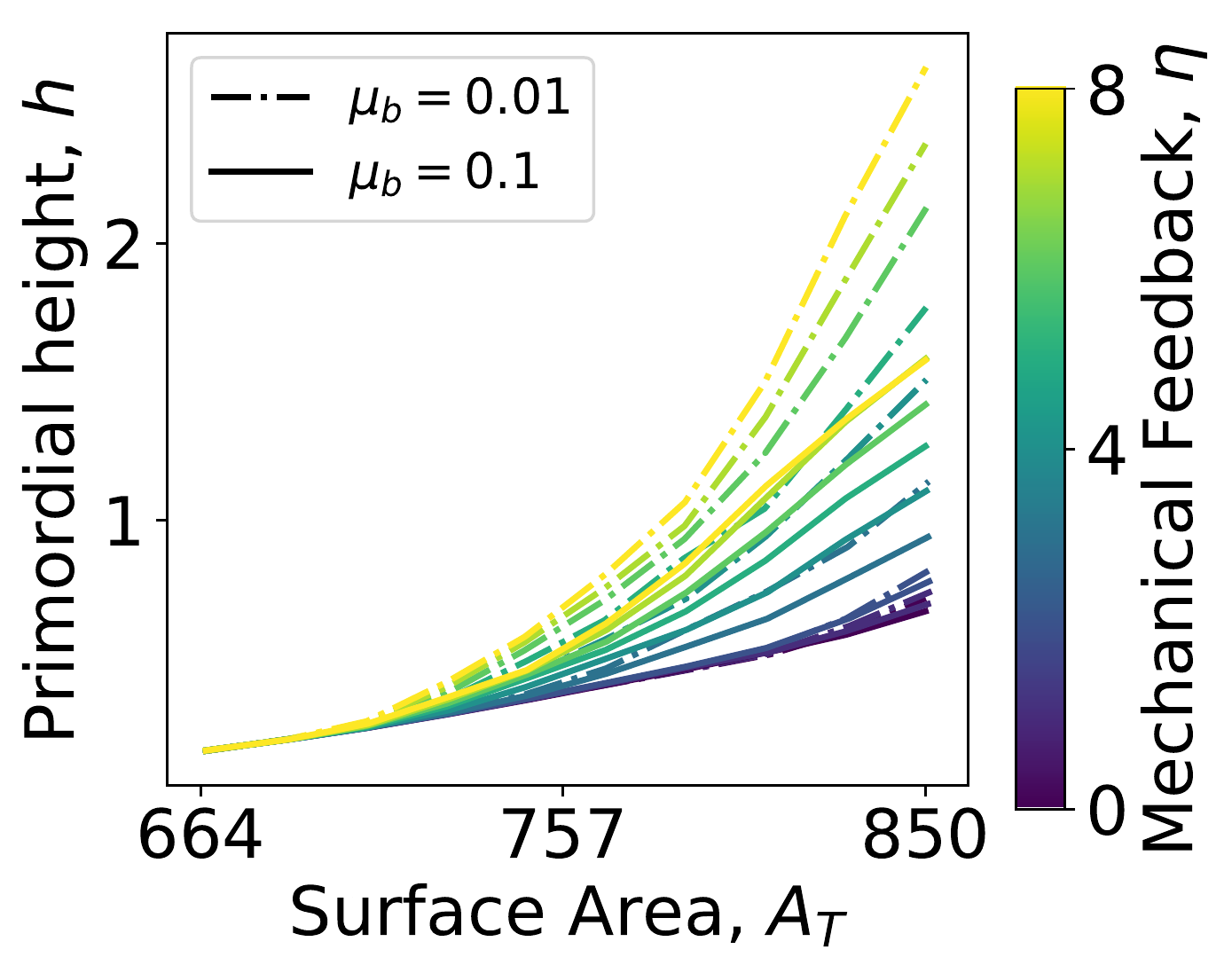}
    \caption{With the change in bending stiffness, the growth curve of primordial height changes but the trend remains the same. By lowering bending stiffness, the growth of primordia becomes faster as the outer bending of primordial cells form the meristematic surface becomes easier.}
    \label{fig:varyBending}
\end{figure}

\begin{figure}[hbt!]
\centering
	\includegraphics[width=0.8\linewidth]{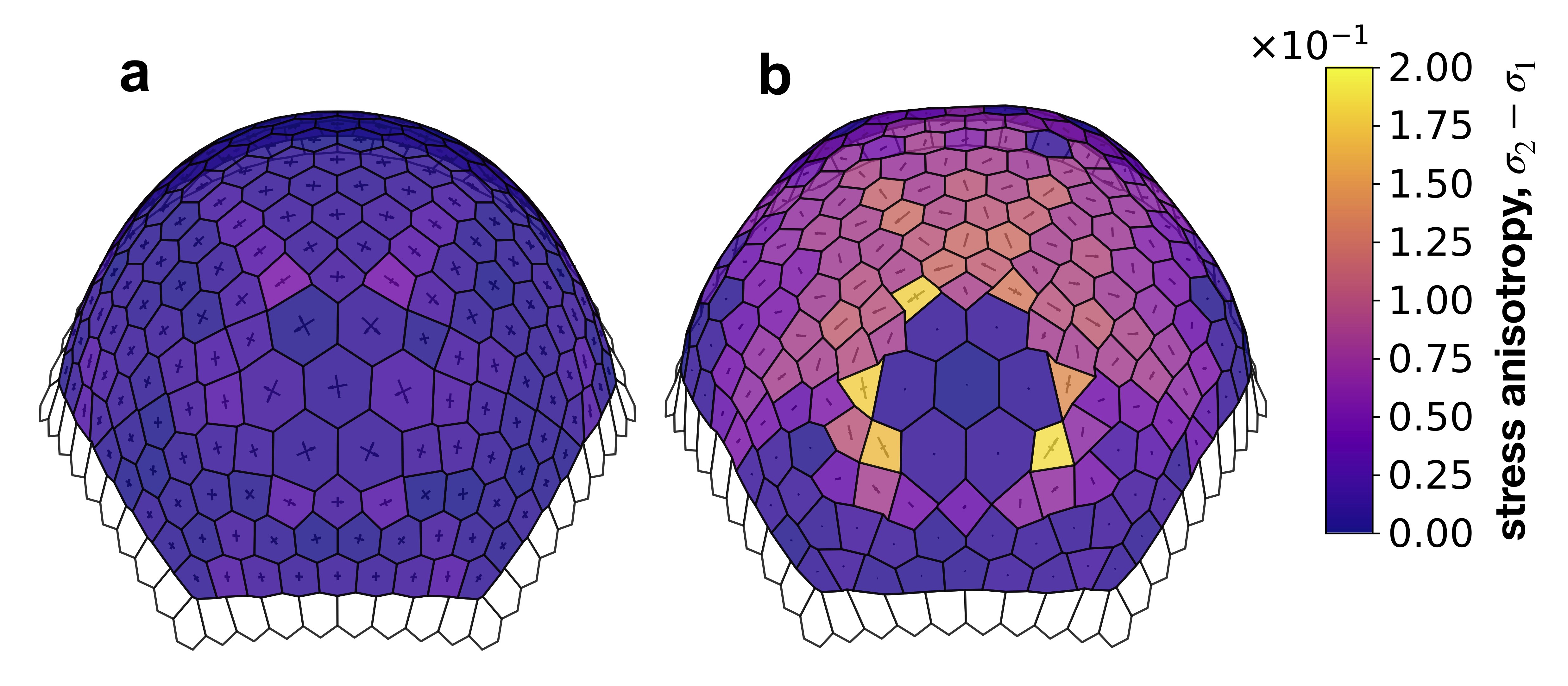}
    \caption{Principal stresses on the cells predicts the along the boundary alignment of microtubules in the boundary cells. (a) and (b) show the tissue of  $A_T = 850$ with stress anisotropy and the principal direction of stresses overlaid on the cells for feedback strength of $\eta = 0$ and $8$ respectively. }
    \label{fig:stressAnisotropyOverlaid}
\end{figure}

\begin{figure}[hbt!]
\centering
	\includegraphics[width=0.6\linewidth]{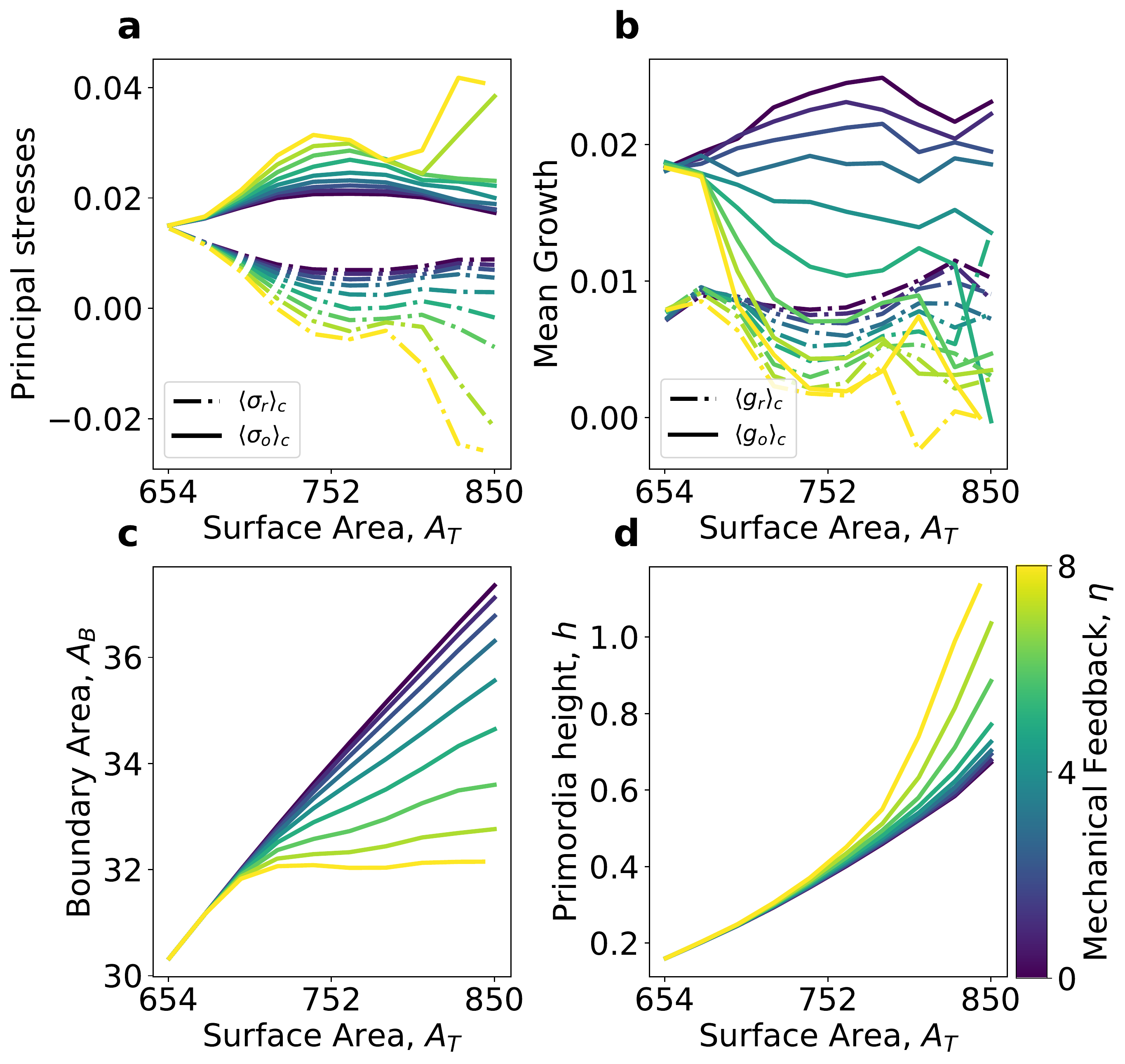}
    \caption{The impact of mechanical feedback was intact with the inclusion of Poisson ratio in the simulation as expected. The figure shows the trend in primordial and boundary growth for Poisson ratio $\nu = 0.375$ and $r_g = 4.8$. (a) and (b) show the radial and orthoradial stresses and growth rates. (c) shows the boundary area around the primordia. The area of the  boundary ceased in growth with feedback but the compression was negligible in this case. (d) The primordial height follow the same trend as before, increasing with higher feedback.}
    \label{fig:varyPoissonRatio}
\end{figure}

\begin{figure}[hbt!]
\centering
	\includegraphics[width=0.8\linewidth]{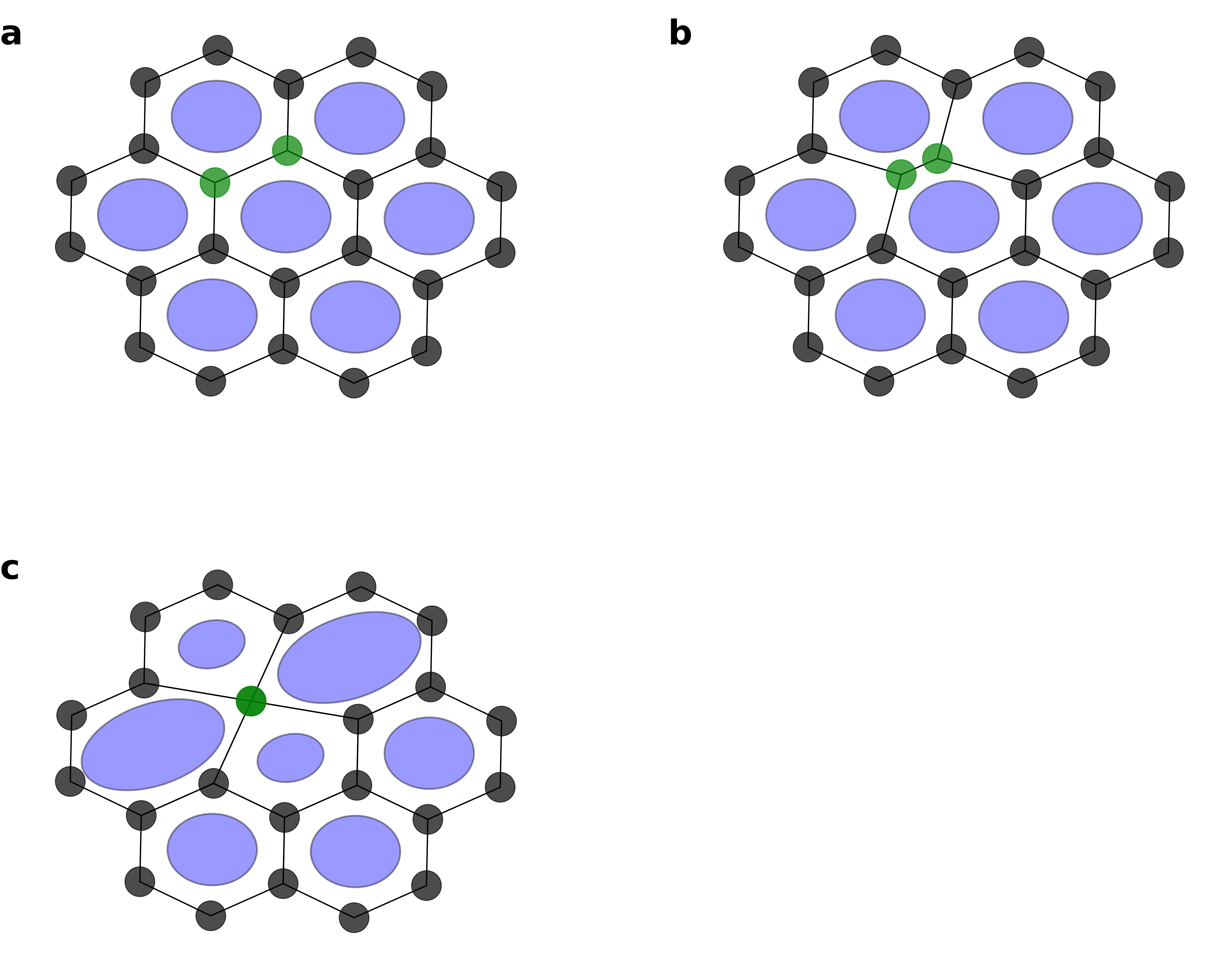}
    \caption{Current cell shape for the cells are robust even under completely shrinking walls. In the plots $a-c$, two vertices (in green) on a 2D cell layer are moving closer until complete overlap and the wall between them shrinks. The cell shape matrix (blue) is mapped as an ellipse, described by eigenvectors and eigenvalues, on the cells. Even with complete shrinking of wall in $c$, the shape matrix is not distorted and represents the orientation and shape of the cells.}
    \label{fig:cellDistorted2d}
\end{figure}

\begin{figure}[hbt!]
\centering
	\includegraphics[width=0.8\linewidth]{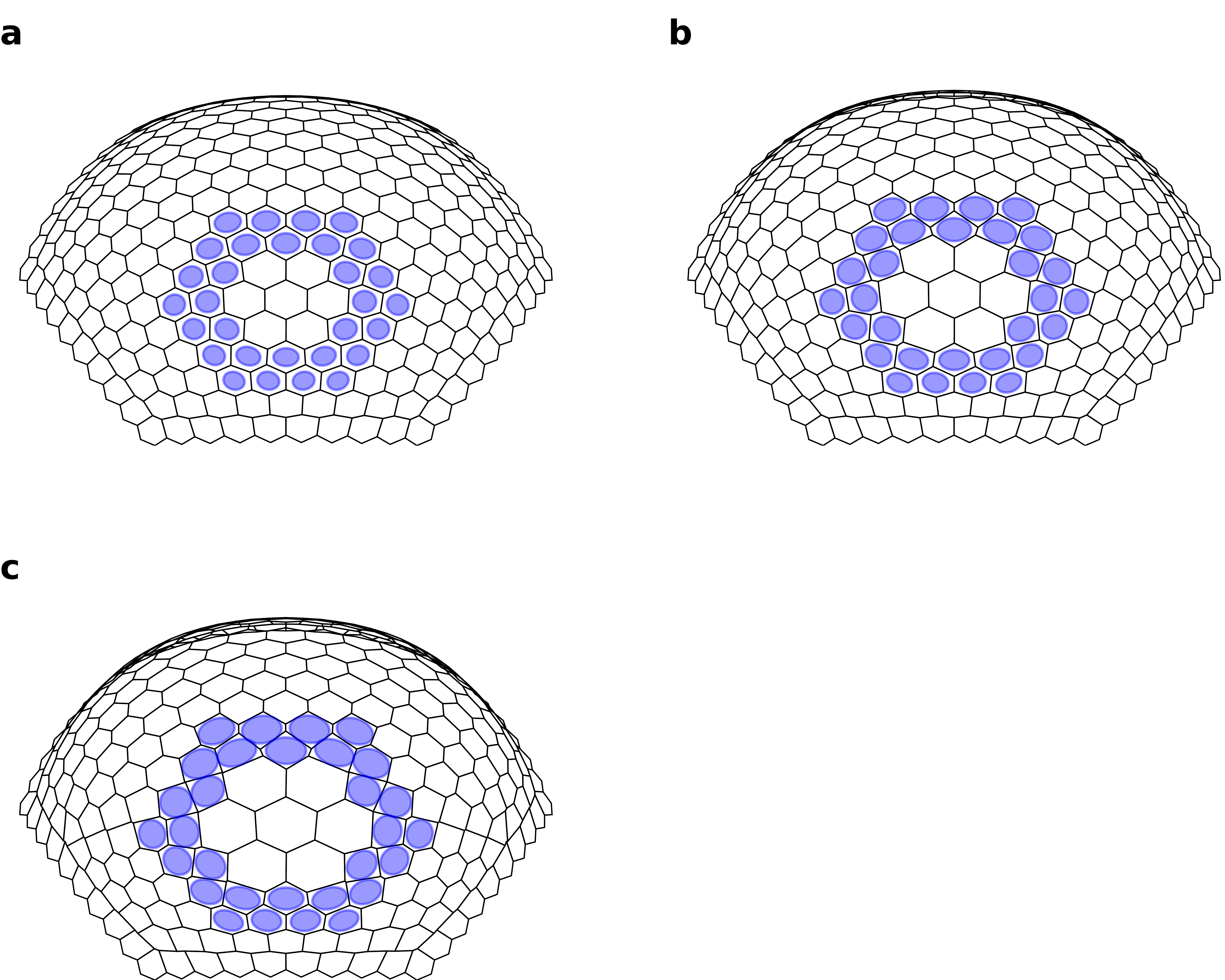}
    \caption{The robustness of cell shape matrix under unnatural distortion of the cell walls and shapes is the same in 3D. $a-c$ show the surface plot of tissue simulation with $r_g = 4.8$ and $\eta = 0$ for $A_T = 720, 785, 850$ respectively. The cells on the boundary face tremendous stress and thus have some unnatural shrinking walls in later stages of simulation ($c$). However, the  cell shape matrices are robust under these distortions and still faithfully represent the actual shape and orientation of the cells. }
    \label{fig:cellDistorted3d}
\end{figure}

We developed a three-dimensional vertex model for plant development to  understand how a primordium, as precursor of aerial organs, can grow out on the shoot apical meristem given the tight connections of cells via plant cell walls. Following the initialization by biochemically triggered local wall softening resulting in higher growth rate in the primordial region we quantify outgrowth dynamics by organ height above tissue level. Taking into account mechanical feedback mediated by cortical microtubules which reinforce cell walls in the direction of higher mechanical stress and promotion of growth in the orthogonal direction, we observe higher and more efficient primordium outgrowth. \\
The cell-based approach of the vertex model for plant tissue developed here ensures the direct coupling of growth of the cells in different directions (expressed with Eq.~\ref{eq:restShapeLockhartGrowthFeedback}). This removes the requirement of additional equations and parameters for modelling the mechanical feedback in the cellular growth.  The artifacts displayed, during later stages of simulations, on the wall shapes of the cells (visible in Fig.~\ref{fig:primiordia-nonprimiordia-ChangeComparision} $b-d$ with shrinking walls in some boundary cells) has little impact on the cell shape computation. The shape matrices defined by second moment of area depend on areal distribution of the cells. They are not distorted with simple artifacts such as shrinking of walls (Fig.~\ref{fig:cellDistorted2d} and \ref{fig:cellDistorted3d}). This is a major advantage for using cell-based vertex model rather than the wall based one which can be highly susceptible to shape distortions. The cell-based shape method also facilitates direct comparison of simulation results with experimental data, which are abundant with mapped out differences in growth at cellular level \cite{kwiatkowska2005flower}. The experimental data at the wall level is still scarce for comparison.
\\
Keeping in mind the robust growth rates in plant tissues, we used the surface area of the meristem as a proxy for age to compare the morphology of the tissue across different parameters. Higher growth rates in the primordial region with respect to the surrounding meristematic tissue were sufficient to trigger organ outgrowth (Fig.~\ref{fig:growthRatioChange} \textit{b}). We found that the absolute values of primordial and meristematic growth rates are irrelevant since the dynamics of primordium formation is dictated by the \emph{ratio} of growth rates between the faster growing primordial cells to the slower growing meristematic cells (Fig.~\ref{fig:growthRationChangeComparision}). However, with mechanical feedback of cell growth on tissue-wide mechanical stresses organ shaping is more efficient.\\
While mechanical feedback does not strongly impact the overall growth of an primordium in area, it directly controls the height of the primordium (Fig.~\ref{fig:feedbackGrowth}). Mechanical feedback can account for the same height with half the growth rate as seen in the following example: the primordia of tissue with growth ratio $r_g = 4.8$ and mechanical feedback of $\eta = 8$ was able to grow to the same height ($h = 1.6$ at $A_T = 850$) as the tissue with twice the growth ratio $r_g = 9.6$ without feedback. Thus, utilizing the CMT-mediated mechanical feedback, plants are able to push out organs from SAM in a faster and efficient manner. \\
We found that the surprising increase in organ height is due to the reorganization of growth and stress on the cells at the boundary of primordium  (Fig.~\ref{fig:feedbackROStress} \textit{a-c}). Boundary cells are under considerable anisotropic stress and this stress anisotropy is further enhanced by mechanical feedback (Fig.~\ref{fig:feedbackROStress} $a$). Larger stresses along the boundary of the growing primordia generated by the mechanical feedback (Fig.~\ref{fig:stressAnisotropyOverlaid}) can also account for the emergence of circumferential alignment of CMTs in the boundary region as have been noted in the experiments \cite{hamant2008developmental}. This implicates the CMT alignment, which follows higher stress, with the reorganization of growth by mechanics in plant cells.\\
The mechanical feedback is observed to slow down boundary cell growth and even ceasing growth for high feedback (Fig.~\ref{fig:feedbackROStress} $b$). In addition, the stiffness of walls in the boundary cells are found to be significantly strengthened in the orthoradial direction and loosened in the radial direction by mechanical feedback (Fig.~\ref{fig:stiffnessReorganisationn8} as compared to  Fig.~\ref{fig:stiffnessReorganisationn0}). Since the primordial area is unaffected by the feedback (Fig.~\ref{fig:feedbackGrowth} \textit{a}), the key role of boundary cells is to act as a stiff and slow growing ring on the tissue surface which pushes out the primordium. An earlier study has noted the effectiveness of a stiff ring-like boundary in the development of primordia \cite{boudon2015computational}. Now, we are able to accredit the emergence of such larger scale pattern in the tissue, without a central organizer, to the mechanical feedback in the cellular growth.\\
The boundary region is even compressed due to the strong stresses from the meristem and primordium in the high feedback regime (Fig.~\ref{fig:feedbackROStress} \textit{c}). The decrease in the area of boundary cells is due to the compressive stresses arising from the primordial development (negative radial stresses seen in Fig.~\ref{fig:feedbackROStress} $b$). As the cells are restricted from shrinking from their rest shape (Eq.~\ref{eq:restShapeLockhartGrowth}), this compression of cells is purely elastic. Similar compression of the boundary cells has been noted in vivo in cells surrounding a growing primordium \cite{kwiatkowska2003growth}. With the introduction of Poisson ratio on the system ($\nu = 0.375$), we found that the boundary compression could be reduced (Fig.~\ref{fig:varyPoissonRatio} $c$). But the boundary growth still remained halted to support higher primordial outgrowth (Fig.~\ref{fig:varyPoissonRatio} $a-d$). We, thus, identified an entirely different mechanism that effectively acts analogous to contractile-ring like dynamics also known to cause shape transformations in animal epithelial tissue \cite{misra2016shape}.\\
Our results also indicate that the saddle shape of the boundary region and the anisotropic shapes of the boundary cells are dependent on the mechanical feedback. Larger negative values of Gaussian curvature can be observed for the case with high mechanical feedback, while for no or low feedback, the boundary has, on average, positive Gaussian curvature (Fig.~\ref{fig:gaussiancurvature}). This suggests that the saddle-like boundary with negative curvatures, as observed in experiments \cite{kwiatkowska2003growth,dumais2002analysis}, can emerge from the growth patterns created by mechanical feedback. Along with this, the shape of the cells themselves in the boundary was found to be progressively anisotropic with mechanical feedback (Fig.~\ref{fig:cellroundness}). Again, suggesting that the tissue wide morphology of the cells can be guided by the reorganised mechanics of the tissue by the stress-based feedback.\\
While a decrease in circumferential strain along with the promotion of axial strains in primordium boundary cells has been suggested to promote primordium outgrowth \cite{oliveri2018regulation}, we here show how such growth dynamics can self-organize due to mechanical feedback. We can therefore finally explain experimental observations of very low or no growth and even compression of cells in the boundary region \cite{reddy2004real,aida2006morphogenesis,louveaux2016cell,kwiatkowska2003growth}. The emerging mechanical patterning can also be suspected as the cause for the separation of meristem and primordium as it mechanically establishes a distinct boundary region on the SAM. \\
Correlating primordial height growth rate and boundary cell stress anisotropy for different values of mechanical feedback (Fig.~\ref{fig:anisotropyScatterPlot}), we observe a clear connection substantiating that boundary cell stress anisotropy increase, proportional to mechanical feedback, is driving primordium outgrowth. Interestingly, for high mechanical feedback stress anisotropy and height growth rate saturate. This suggests that the gain in primordial growth flattens out in the high feedback regime and there could be an optimal level of mechanical feedback for efficient growth in plants clarifying previous model observations \cite{alim2012regulatory}.\\
Future investigations can be targeted to explore the influence of other biological processes, such as cell divisions, on the growth of the primordia. Preliminary results from our simulations show the impact of mechanical feedback to remain intact with cell division (Fig.~\ref{fig:cellDivision}). However, there are strong patterns of cell division observed during primordial development. Primordial cells are known to exhibit higher rates of divisions compared to the meristematic cells \cite{laufs1998cellular} and the divisions in boundary cells have been suggested to orient following the stress \cite{louveaux2016cell}. The study of such divisions in primordial growth can elucidate the intricate role of division in morphological development and also help understand the preferential direction and timing of the cells for division.\\
The expansion of the model from two-dimensional surface to a full description in three-dimension can also be considered for the future work. Modulating the bending stiffness for the cells, which represents the stiffness of the anticlinal walls, do show an impact on the magnitude of primordial growth (Fig.~\ref{fig:varyBending}). A feedback of stresses with stiffness in the anticlinal direction can also be examined to further understand the over all regulation of plant cell growth by tissue mechanics.\\
Taken together, our key insight is that mechanical feedback
reorganizes cell growth by two distinct mechanisms. First,
feedback directly influences cell growth by modulation of
wall properties. Second, feedback changes the stress patterns on cells, thereby self-amplifies and propagates growth
anisotropies that then indirectly influence cell growth again.
This twofold mechanism allows plant tissue to initiate organ outgrowth efficiently by modifying their growth pattern
through stress feedback and rather than the amplifying growth
rates at the expense of cell material.
\section*{Author Contributions}
J.K., J-D.J. and K.A. designed research. J.K. performed the research. J.K., J-D.J. and K.A. wrote the article.
\section*{Acknowledgments}
This work was in part supported by the Max Planck Society and the
Deutsche Forschungsgemeinschaft via grants SFB-937/A19 and DFG-FOR 2581.

\clearpage


\clearpage
\onecolumn
\beginsupplement
\begin{center}
      \Large\textbf{Feedback from tissue mechanics self-organizes efficient outgrowth of plant organ}\\[0.5em]
      \Large\textbf{Supplementary Material}\\[0.5em]
      \large{J. Khadka, J-D. Julien and K. Alim}
\end{center}
\section{Simulation parameters}
The parameters used for the simulation of growth of SAM and primordia are given in Table~\ref{tab:simulationparameters}. The Lamé's first parameter $\mu$, bending stiffness $\mu_b$ and pressure $P$ were kept constant through all simulations.  The noise in the growth, given by $\gamma$, was defined as a uniform noise in the range $[-0.1,0.1]$, thus simulating $10\%$ deviation of growth rates from their initial values. The analysis for feedback were based on the analysis of simulations with a  growth ratio $r_g = 4.8$, and the same results holds true for varying growth rate, checked by increasing the growth ratio to $r_g = 9.4$.   
\renewcommand{\arraystretch}{1.2}
\begin{table}[h]
    \centering 
    \begin{tabular}{@{}crcrcrcrccrcc@{}}\toprule
    \phantom{abc}&\phantom{abc}&\phantom{abc}&\phantom{abc}&\phantom{abc}&\phantom{abc}&\phantom{abc}&\phantom{abc}&\multicolumn{2}{c}{$r_g = 9.4$}& \phantom{abc}& \multicolumn{2}{c}{$r_g = 4.8$}\\
    \cmidrule{9-10} \cmidrule{12-13} $\mu$ & \phantom{abc}&$\mu_b$& \phantom{abc}&$P$& \phantom{abc}& $\gamma$&& $\kappa_f$ & $\kappa_s$ & & $\kappa_f$ & $\kappa_s$ \\\midrule
    $1.0$ & &$0.1$& &$0.0126$& & $\mathcal{U}(-0.1,0.1)$&&$0.5$&$0.05$&& $0.5$&$0.1$\\
    \bottomrule
    \end{tabular}\vspace{0.2cm}
    \caption{Parameter values used for simulation of primordial growth using three-dimensional vertex model. $\mathcal{U}(-0.1,0.1)$ is an uniform random number taken in the interval of $[-0.1,0.1]$.}
    \label{tab:simulationparameters}
\end{table}
\section{Positive semi-definite criteria on cellular growth}
To ensure the cellular growth is plastic, the rest shapes of the cells are prevented from shrinking through a positive semi-definite criteria on the growth. To this end, the operation $(.)_+$ on a symmetric second order tensor $T$ with rank decomposition given in Eq.~\ref{eq:TRankDecomposition}, with  $\lambda_n$ and $t_n$ as eigenvalues and eigenvectors, can be written as Eq.~\ref{eq:positiveSemiDefiniteCriteria}.
\begin{align}
	T &= \sum^d_{n = 1} \lambda_n t_n \otimes t_n
    \label{eq:TRankDecomposition}
    \\
    (T)_+ &= \sum^d_{n = 1} max(\lambda_n ,0) t_n\otimes t_n 
    \label{eq:positiveSemiDefiniteCriteria}
\end{align}
\section{Primordial height}
\begin{figure}[hbt!]
\centering
		\includegraphics[width=0.4\linewidth]{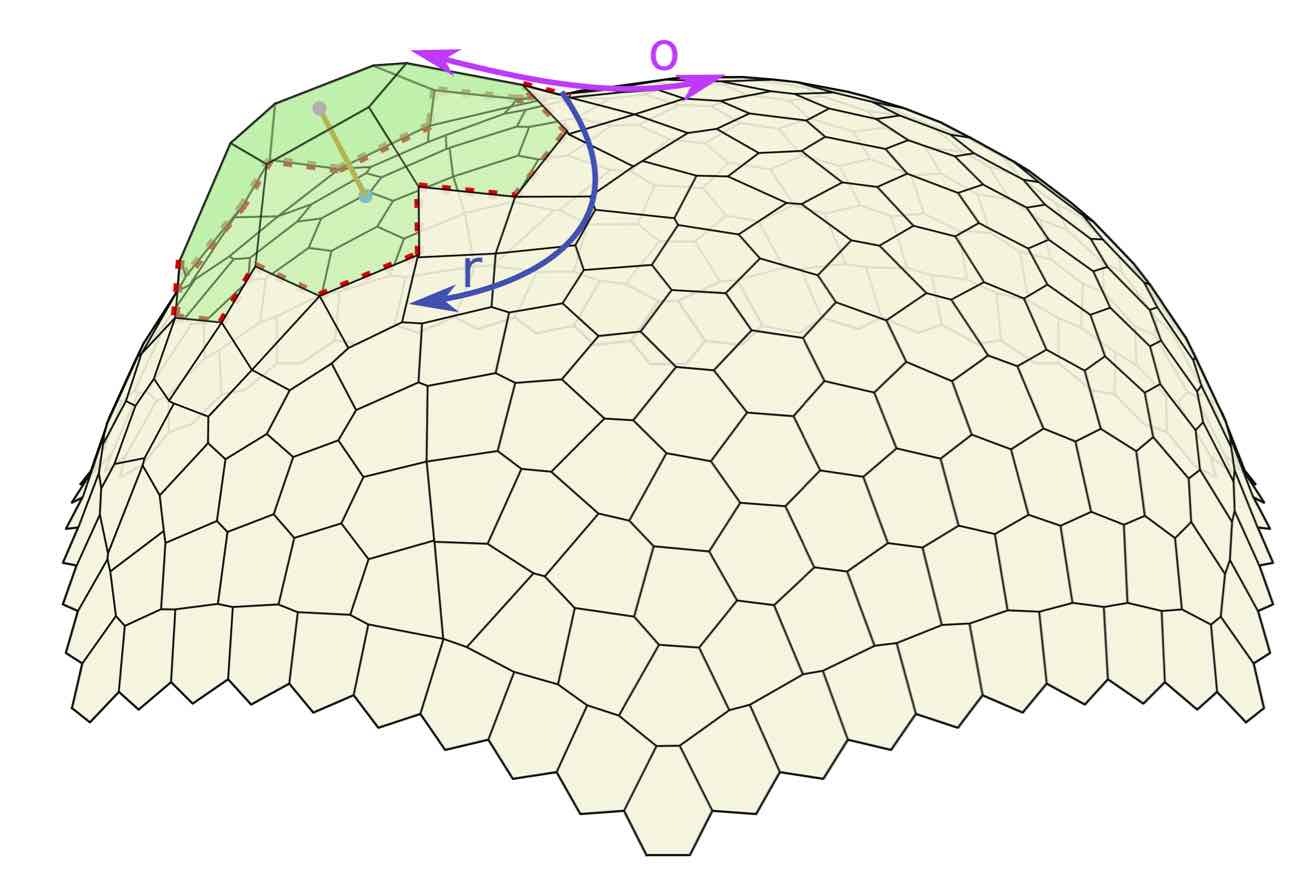}
	\caption{The height of the primordium (green) above the shoot apical meristem is shown in orange. The red dotted line shows the primordium boundary. The purple and the blue arrow show the orthoradial and radial direction for stress and growth respectively. }
	\label{fig:growthHeightMeasurement}
\end{figure}

The primordial height is taken as the distance form the base of primordia to the top (shown by orange line in Fig.~\ref{fig:growthHeightMeasurement}). The base is taken as the average position of boundary vertices and the top position as the centroid of the cell at the top of the primordia. 
\section{Classification of cells on the SAM}
The cells on the SAM are classified as the primordia, meristem and boundary cells for the quantitative analysis. Primordia cells are distinguished, as the cells that grow fast and result in the bulging primordia. The boundary cells are defined as the two layers of cells around the primordia that are shown to have significantly different properties. They are positioned between the primordia and the rest of the SAM and thus face highly anisotropic mechanical stresses. The rest of the cells on the tissue are identified as meristem cells. These cells are far away from the primordia to not face significant mechanical changes induced by primordial growth. 

\section{Bending energy of the tissue}
The tissue in the vertex model is a discrete mesh of vertices, which poses challenges for the computation of $H$. We adapted the discretization developed by Meyers et al.~to compute Eq.~\ref{eq:helfrichBendingComplete} for the tissue \cite{meyer2003visualization,guckenberger2016bending}.\\
The local mean curvature around a point on the surface can be rewritten as 
\begin{equation}
	H(\vec{x}_i) = \frac{1}{2} (\Delta_S \vec{x}_i) \cdot \vec{n}(\vec{x}_i), \qquad \vec{x} \in S \text{.}
	\label{eq:meanCurvatureExpanded}
\end{equation}
$\vec{n}(\vec{x}_i)$ is the normal vector at point $\vec{x}_i$. The operator $\Delta_S$ is the Laplace-Beltrami operator for the surface $S$, and is expressed as, 
\begin{equation}
	\Delta_S = \nabla^S \cdot \nabla^S \text{,}
	\label{eq:laplaceBeltrami}
\end{equation}
with $\nabla_S$ as the gradient of the surface. The mean curvature $H$ can be obtained from the operator $\Delta_S$ by rewriting Eq.~\ref{eq:meanCurvatureExpanded} as
\begin{equation}
	H = \frac{1}{2}\Vert{\Delta_S\vec{x}}\Vert \text{.}
	\label{eq:meancurvatureLBO}
\end{equation}
\begin{figure}[hbt!]
	\centering\includegraphics[width=0.4\linewidth,trim={2cm 0cm 2cm 0cm},clip]{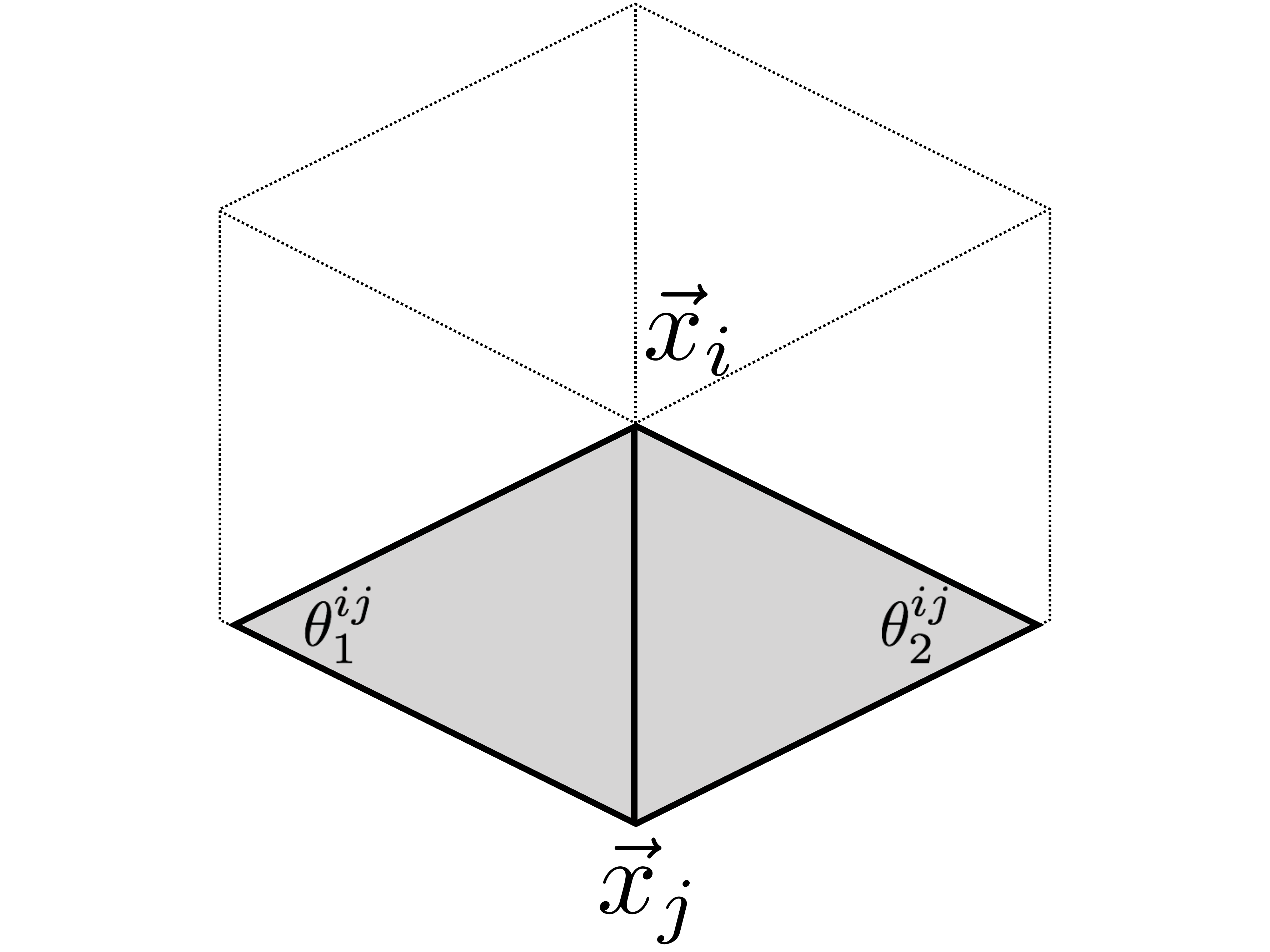}
	\caption[1-ring neighbours of a vertex $\vec{x}_i$.]{For a vertex $\vec{x}_i$,  1-ring neighbours are all the vertices that are joined by an edge to $\vec{x}_i$ in the mesh and $\theta_1$ and $\theta_2$ are opposite angles to the edge joining vertex $\vec{x}_i$ and its neighbour.}
	\label{fig:1-ring-neighbours}
\end{figure}

\begin{figure}[hbt!]
	\centering
	\subfigimg[height=3.2cm,trim={1cm 1cm 0 4cm},clip]{a}{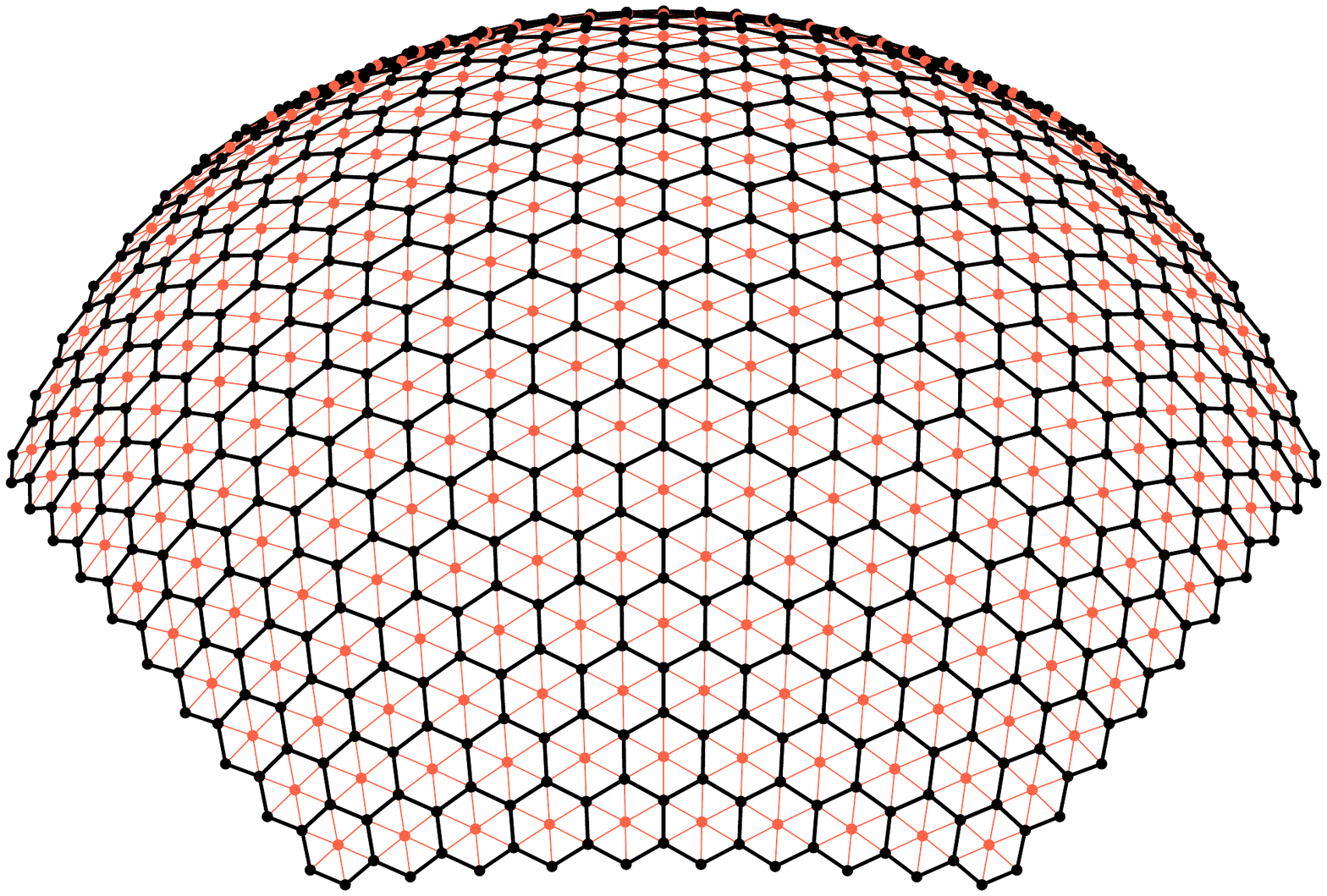}
	\subfigimg[height=3.2cm,trim={1cm 1cm 0 4cm},clip]{b}{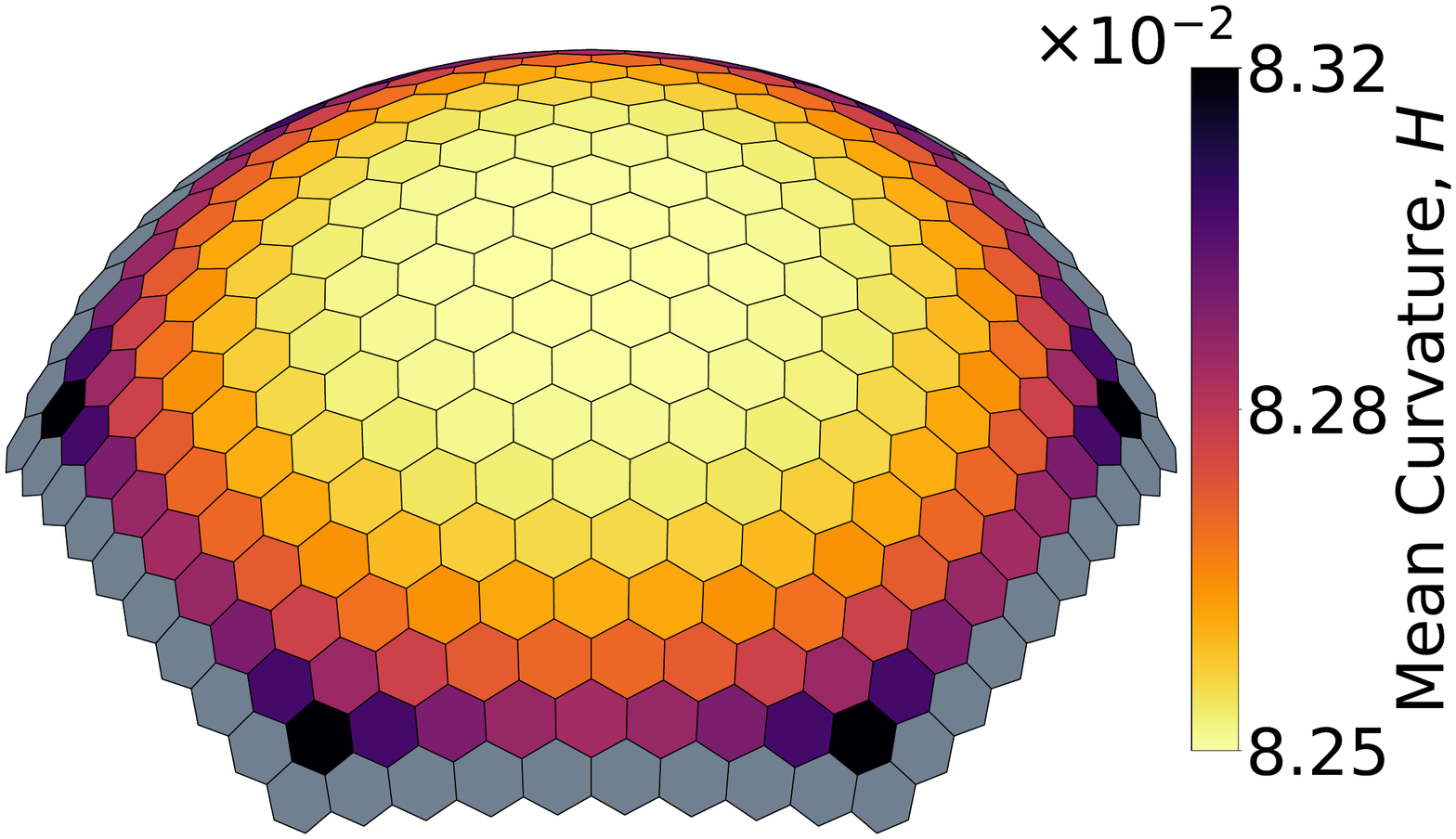}
	\caption[Triangular discretisation of polygonal mesh.]{(a) The hexagonal cells of the tissue (in black) is further discretised using the centroids (red points) of the cells. (b) Mean curvature of the initial dome like tissue. For each cell, the mean curvature is calculated as an average of the curvature at its vertices and centroid. The boundary cells (in grey) are excluded in the plot as the boundary vertices have significantly high curvature. This artifact can be ignored since the boundary is fixed in its position. The slight variation in the mean curvature on the dome cells is the result of the hexagonal discretisation of hemispherical surface.}
	\label{fig:meshTriangulation}
\end{figure}

We can write a discretisation of Laplace-Beltrami operator on a triangulated mesh by considering directly connected neighbours for each vertex $\vec{x_i}$. We call these connected neighbours as 1-ring neighbours for a vertex. The discretisation of the operator is then obtained by a contour integral around 1-ring neighbouring vertices of a vertex $\vec{x}_{i}$ (Fig.~\ref{fig:1-ring-neighbours}) as \cite{meyer2003visualization}
\begin{equation}
	\Delta_S w(\vec{x}_{i})\approx \frac{\sum_{ji} (cot\theta_1^{ij}+cot\theta_2^{ij})(w(\vec{x}_{i})-w(\vec{x}_{j}))}{2A^{i}_{mixed}} \text{.}
	\label{eq:LBOdiscretisedArbitrary}
\end{equation}
$\theta_1^{ij}$ and $\theta_2^{ij}$ are the angles opposite to the edge joining vertex $\vec{x}_i$ and $\vec{x}_j$ in the triangular mesh (Fig.~\ref{fig:1-ring-neighbours}). $w$ is an arbitrary two-times continuously differentiable function on $S$, which can be taken as the position $\vec{x}_{i}$ itself, and the above equation can be expressed as 
\begin{equation}
	\Delta_S \vec{x}_{i}\approx \frac{\sum_{ji} (cot\theta_1^{ij}+cot\theta_2^{ij})(\vec{x}_{i}-\vec{x}_{j})}{2A^{i}_{mixed}}
	\label{eq:LBOdiscretised} \text{.}
\end{equation}
The summation is over all 1-ring neighbouring vertices of vertex $\vec{x}_{i}$. $A^{i}_{mixed}$ is the mixed area for the vertex $\vec{x}_{i}$. It is calculated as described in Algorithm \ref{alg:area_mixed} to insure the $A_{mixed}$ for all vertex will tile the surface \cite{meyer2003visualization}. Either Voronoi area of a vertex or a fraction of triangular area ($area(T)$) from the neighbourhood of the vertex is summed depending on the condition defined in Algorithm \ref{alg:area_mixed} to calculate the mixed area. The Voronoi area for a vertex $\vec{x}_i$ can be calculated as
\begin{equation}
	A_{Voronoi} = \frac{1}{8} \sum_{j}(cot \theta^{ij}_1 + cot \theta^{ij}_2)\Vert{\vec{x}_i-\vec{x}_j}\Vert^2  \text{,}
	\label{fig:vornoiarea}
\end{equation}
where the sum is again around the 1-ring neighbours of the vertex.\\
\begin{algorithm}[hbt!]
    \SetAlgoVlined
    $A_{mixed} = 0 $\\
    \For{each triangle $T$ from the 1-ring neighborhood of $\vec{x}$}{
    \uIf(\tcp*[h]{Voronoi safe}){$T$ is non-obtuse}{
    \tcp{Add Voronoi formula}
    $A_{mixed} +=$ Voronoi region of $\vec{x}$  in $T$}
    \Else(\tcp*[h]{ Voronoi inappropriate}){
    \tcp{Add either $area(T)/4$ or $area(T)/2$}
    \eIf{the angle of $T$ at $\vec{x}$ is obtuse}{ 
    $A_{mixed} += area(T)/2$}
    {
    $A_{Mixed} += area(T)/4$}
    }
    }
    \caption{Algorithm to calculate $A_{mixed}$ on an arbitrary mesh \cite{meyer2003visualization}}
    \label{alg:area_mixed}
\end{algorithm}\\\noindent
Since the tissue surface is tiled with hexagonal cells, we triangulate the hexagonal lattice for the calculation of mean curvature by using the centroid of the cells as shown in Fig.~\ref{fig:meshTriangulation} $a$. The complete discretised form of Eq.~\ref{eq:helfrichBendingComplete} on the triangulated tissue can then be expressed as
\begin{equation}
	U_{bending} = 2 \mu_b \sum_{\vec{v}_T} (H(\vec{v}_T)-H_0(\vec{v}_T))^2\text{,}
	\label{eq:discreteHelfrichBending}
\end{equation}
where $\vec{v}_T$ includes all the nodes of triangulated mesh, \textit{i.e.} all the vertices and the centroid of the cells. \\
Similarly, a discretised expression for Gaussian curvature $K$ at a vertex on the triangulated mesh can be written as
\begin{equation}
	K(\vec{x}_i) =\frac{1}{A_{mixed}} \left( 2\pi - \sum_f \theta_f \right) \text{.}
	\label{eq:gaussiancurvature}
\end{equation}
The summation is over 1-ring neighbouring triangulated faces $f$ of vertex $\vec{x}_i$ (Fig.~\ref{fig:1-ring-neighbours} and \ref{fig:meshTriangulation} $a$) and $\theta_f$ is angle at vertex $\vec{x}_i$ in triangle $f$. 
\section{Calculation of growth ratio and uniform dynamics of primordia}
\begin{figure}[hbt!]
\centering
	\centering
	\includegraphics[width=0.8\linewidth]{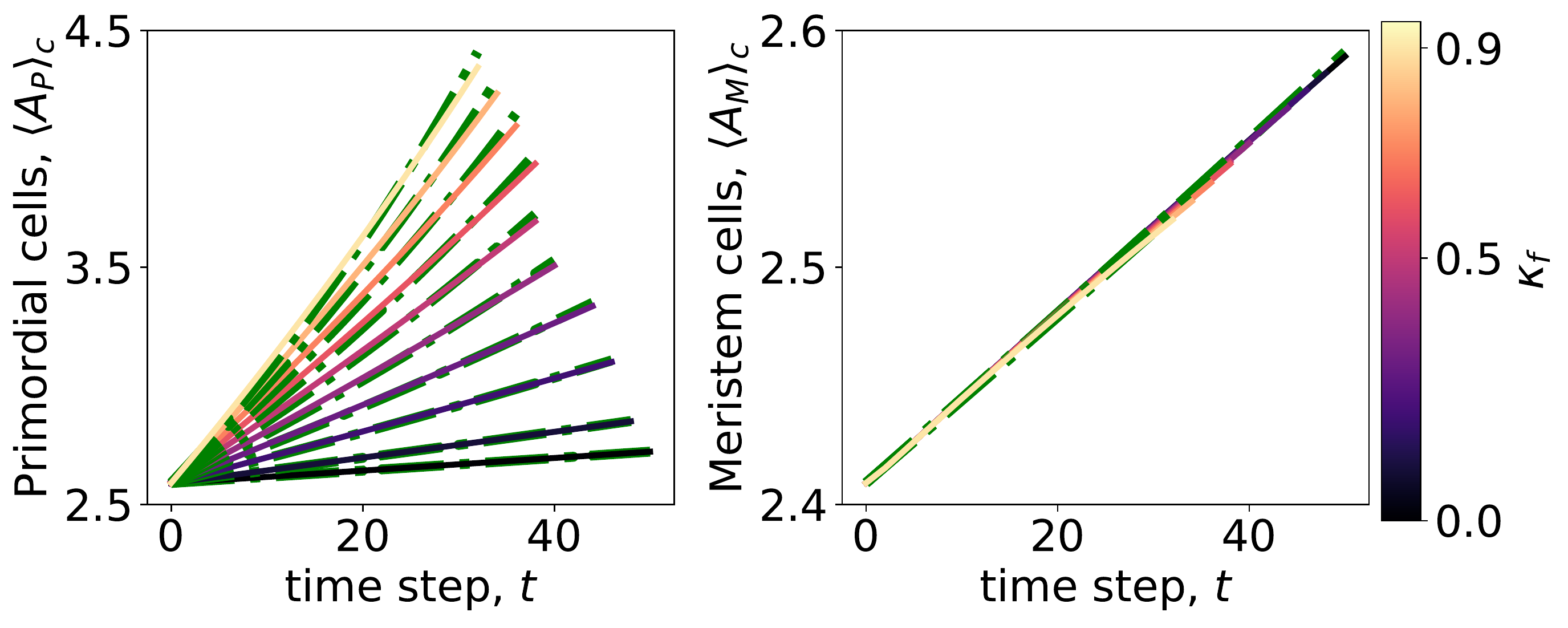}
	\caption{The growth curve obtained from the simulation are fitted with exponential growth curve to get the actual growth rate of cells ($\kappa^*$). (a) and (b) show the averaged cellular area for primordial cells $\langle A_P \rangle_c$ and for meristem cells $\langle A_M \rangle_c$ respectively. The meristem growth rates are kept constant ($\kappa_s = 0.05$), while the primordia l growth rates ($\kappa_f$) are varied as shown by the color map. The solid line with color map show the averaged cellular area obtained from simulation and dotted green line show the fitted exponential curves. (a) Increasing $\kappa_f$ leads to larger averaged area for primordial cells. (b) The growth of meristem cells is unchanged as $\kappa_s$ is kept constant.}
	\label{fig:varyfkexponentialfit}
\end{figure}

The input growth rate of cells ($\kappa$ in Eq.~\ref{eq:restShapeLockhartGrowth}) define the growth of rest cell shape. However, the actual cell shape on the tissue is not only dependent on the rest cell shape but the shape of surrounding cells and curvature and volume of the tissue also effect the final shape of the cell. To measure the actual growth rates of the cells, we take the areal growth curve of a cell obtained from the simulation and fit an exponential growth ($A = A_0e^{\kappa^* t}$) on it. Fig.~\ref{fig:varyfkexponentialfit} shows the areal growth curve and exponential fit for varying growth rates of primordial cells and meristem cells. Table~\ref{tab:growthratiotable} shows the comparison between the input growth rates to the calculated growth rates from the growth curves. \\
\\

\begin{figure}[hbt!]
\centering
	\centering
	\includegraphics[width=0.7\linewidth]{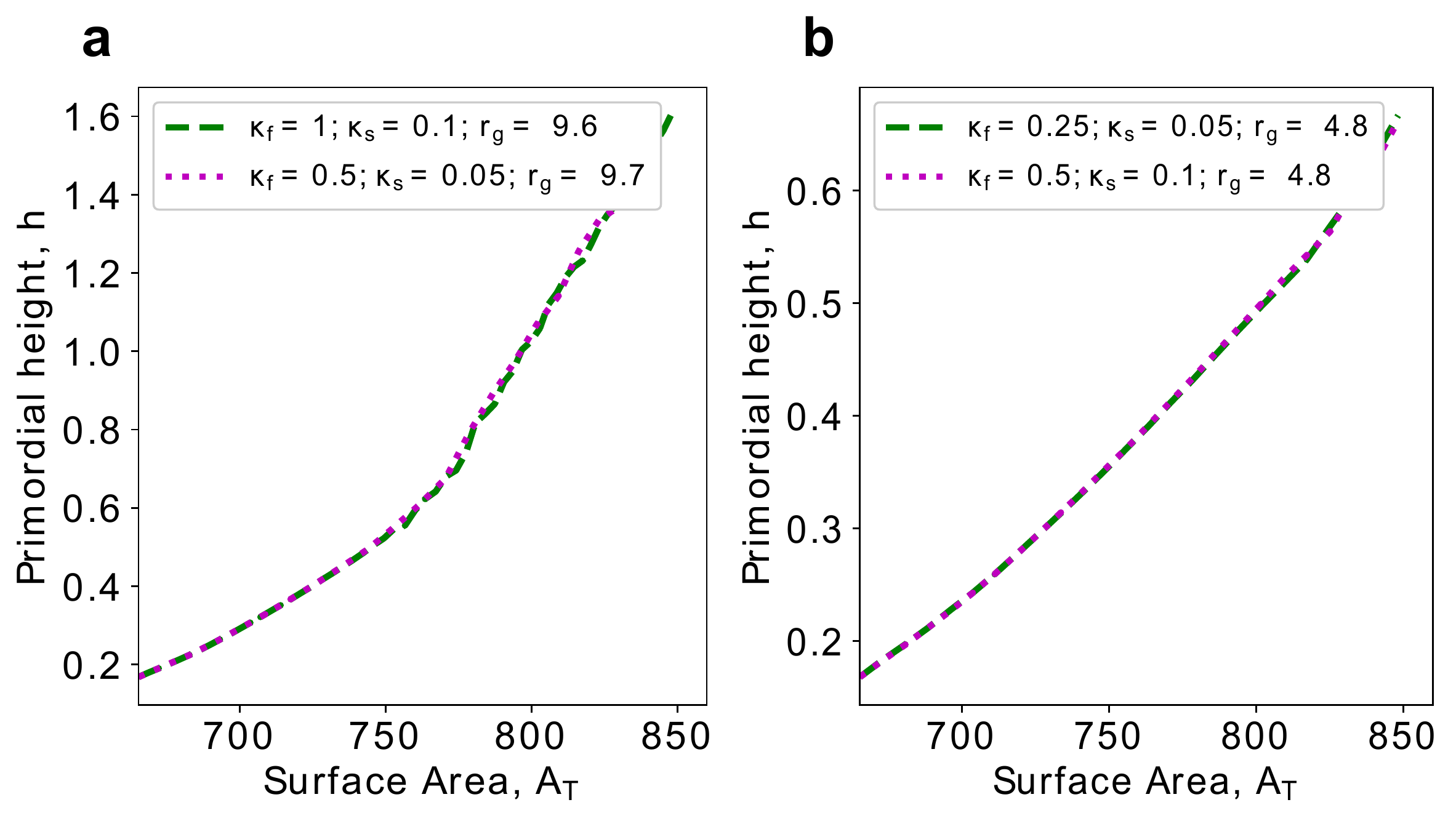}
	\caption{The same growth ratio describes same primordial outgrowth dynamics with respect to overall tissue growth. The comparison between two different sets of growth rates keeping the growth ratio fixed shows that the outgrowth dynamics is dictated by the growth ratio. (a) Simulation with growth ratio $r_g=9.6$. (b) Simulation with growth ratio $r_g=4.8$ }
	\label{fig:growthRationChangeComparision}
\end{figure}

\begin{table}[h]
\centering 
{
    \begin{threeparttable}
    \captionsetup{labelfont={},font={small},justification=raggedright,
              singlelinecheck=false}
    \begin{tabular}{@{}crcrcrcrcrc@{}}\toprule
    \toprule
    $\kappa_s$ && $\kappa_f$ &&$\kappa^*_s$ && $\kappa^*_f$&& $\kappa_f/\kappa_s$ && $r_g = \kappa^*_f/\kappa^*_s$\\\hline
    0.05 &&  0.0    && 0.00092     &&  0.00099    && 0.0 &&   1.1 \\
 0.05 &&  0.05    &&  0.00091    &&  0.00099    && 1.0 &&   1.1\\
 0.05 &&  0.1    &&   0.00091   &&  0.00190     && 2.0 &&   2.1\\
 0.05 &&  0.15    &&  0.00091    &&  0.00275     && 3.0 &&  3.0\\
 0.05 &&  0.2    &&   0.00091   &&  0.00356     && 4.0 &&   3.9\\
 0.05 &&  0.25    &&  0.00090   &&  0.00434    && 5.0 &&    4.8\\
 0.05 &&  0.3    &&  0.00090   &&  0.00505     && 6.0 &&    5.6\\
 0.05 &&  0.35    &&  0.00089   &&  0.00564    && 7.0 &&    6.4\\
 0.05 &&  0.4    &&  0.00087   &&  0.00612    && 8.0 &&    7.0\\
 0.05 &&  0.45    && 0.00086    &&  0.00651    && 9.0 &&    7.6\\
 0.05 &&  0.5    &&   0.00085   &&  0.0068     && 10.0 &&  8.0 \\
 0.05 &&  0.55    &&   0.00084  &&  0.0071     && 11.0 &&   8.4\\
 0.05 && 0.6    && 0.00083  &&  0.00726   && 12.0 &&   8.7\\
 0.05 && 0.65    &&   0.00082  &&  0.00745     && 13.0 &&   9.1\\
 0.05 && 0.7    &&   0.00082    &&  0.00762    && 14.0 &&  9.3\\
 0.05 && 0.75    &&   0.00081   &&  0.00776    && 15.0 &&   9.6\\
 0.05 && 0.8    &&  0.00080   &&  0.00791   && 16.0 &&    9.9\\
 0.05 && 0.85    &&  0.00079     &&  0.00794    && 17.0 &&  10.0\\
 0.05 && 0.9    &&   0.00079   &&  0.00804     && 18.0 &&   10.2\\
 0.05 && 0.95    &&    0.00078   &&  0.00809    && 19.0 &&  10.3\\
    \bottomrule
    \end{tabular}
    \end{threeparttable}
    \vspace{0.2cm}
    \caption{The measurement of growth rates ($\kappa^*_s$ and $\kappa^*_f$) from the simulation as respect to the input growth rates ($\kappa_s$ and $\kappa_f$) and the resulting growth ratio.}
    \label{tab:growthratiotable}
}
\end{table}
The growth ratio obtained from the fitted growth rates of primordial and meristem (as shown in Fig.~\ref{fig:varyfkexponentialfit}) strongly dictate the growth of primordia. The dynamics of primordial growth with respect to the overall tissue growth remains unchanged with varying individual growth rates ($\kappa_f$ and $\kappa_s$) but keeping the growth ratio ($r_g$) constant (Fig.~\ref{fig:growthRationChangeComparision}). We utilize the total surface area of the tissue as a proxy for time as the overall growth rates of the meristem is robust in biology. The total surface area of the tissue includes the primordial area and  with increasing mechanical feedback, the primordial area growth remains stable with tissue surface area growth (Fig.~\ref{fig:feedbackGrowth} $a$). The primordial height, however, is significantly larger with larger mechanical feedback (Fig.~\ref{fig:feedbackGrowth} $b$). Similar is not true for modulating the growth ratio by modulating the growth rates of primordial and meristematic cells. With higher growth ratio as a tissue grows larger by increasing its surface area, a gain in primordial height is observed but it is together with significant increase in the primordial area (Fig.\ref{fig:growthRatioChange} $a,b$). On Fig.~\ref{fig:growthRationChangeComparision}, thus, it is remarkable that changing the growth rates of primordia and meristem while keeping the growth ratio constant does not have any bearing on the height growth dynamics with respect to tissue surface growth. Even though as observed in Fig.~\ref{fig:growthRatioChange}, altering one or the other has a significant impact on the primordial height and the primordia area.


\section{Reorganised stiffness of cells} \label{section:stiffnessmodulation}
\begin{figure}[hbt!]
\centering
	\centering
	\includegraphics[width=\linewidth]{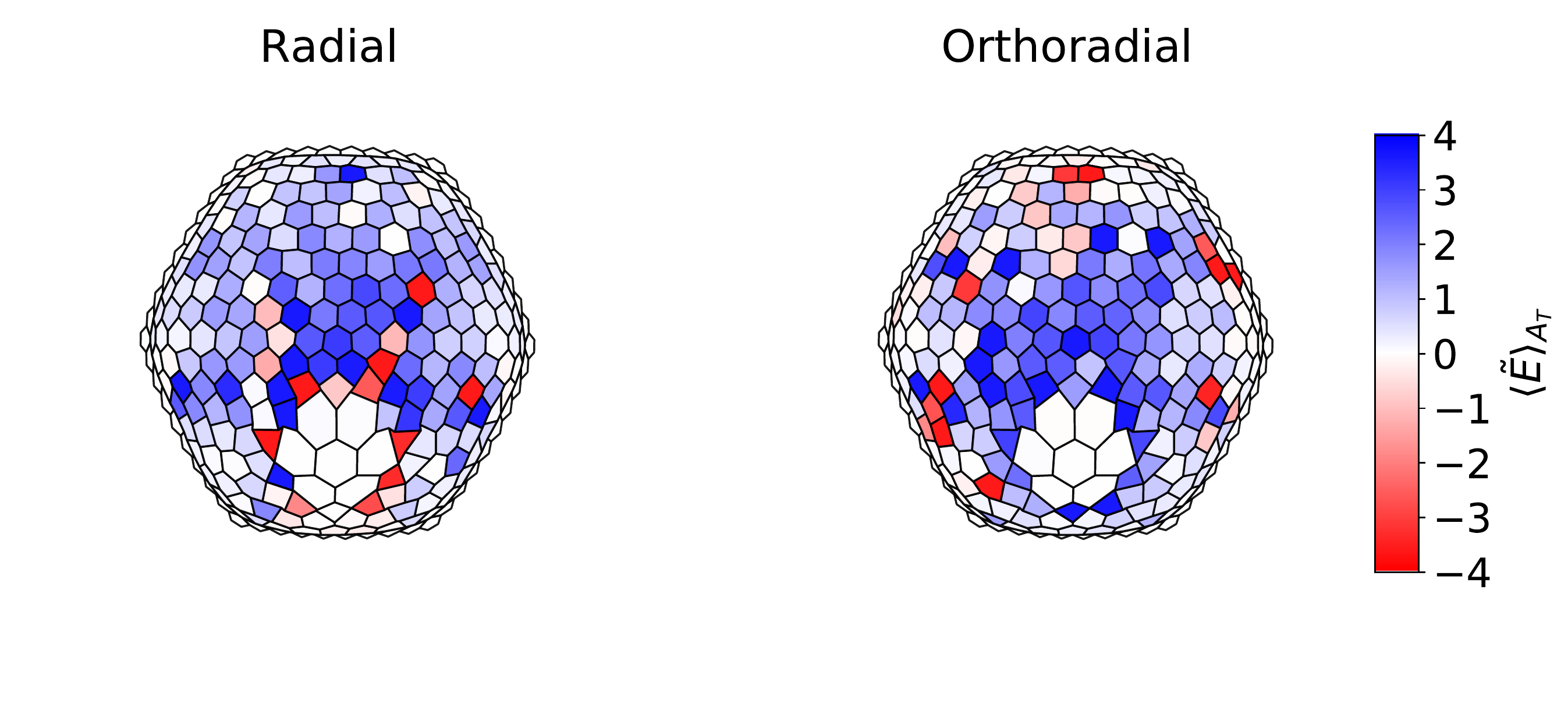}
	\caption{The reorganisation of cellular stiffness by the mechanical feedback plotted by averaging over the total tissue growth ($\langle\tilde{E}\rangle_{A_T}$) for the case with mechanical feedback $\eta = 8$.}
	\label{fig:stiffnessReorganisationn8}
\end{figure}

\begin{figure}[hbt!]
\centering
	\centering
	\includegraphics[width=\linewidth]{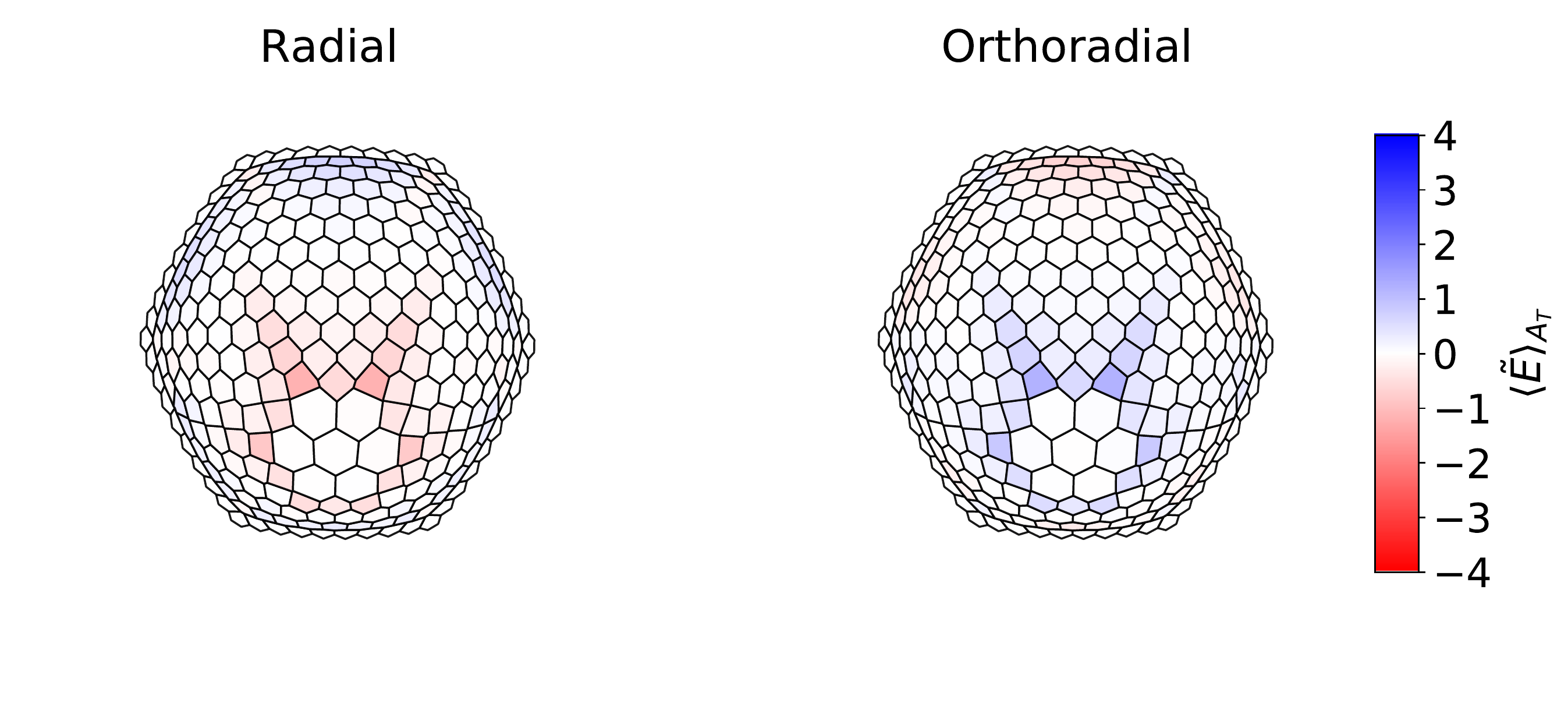}
	\caption{The reorganisation of cellular stiffness by the mechanical feedback plotted by averaging over the total tissue growth ($\langle\tilde{E}\rangle_{A_T}$) for the case without mechanical feedback $\eta = 0$.}
	\label{fig:stiffnessReorganisationn0}
\end{figure}

We can define the relative changes of stiffness for the cells from mechanical feedback by looking at the growth equation for the shape matrix of the cells. The growth equation without the mechanical feedback for the cells is 
\begin{equation}
    \frac{dM^0_c}{dt} = \kappa(1+\gamma) (M_c-M^0_c) \text{.}
\end{equation}
We can express the growth terms in the principal directions, which is following the principal directions of deformation ($M_c - M_c^0$), 
\begin{equation}
    \kappa(1+\gamma) (M_c-M^0_c) = \sum_i \lambda^i_d v^i_d v^{iT}_d\text{.}
\end{equation}
The right hand side represents the eigen decomposition of the matrix. 
With the mechanical feedback, or stress coupling, on the growth of the shape matrix, the growth equation is
\begin{equation}
    \frac{dM^0_c}{dt} =  \kappa(1+\gamma) (M_c-M^0_c)-\frac{\eta}{2} \Big(D_c(M_c-M^0_c) +
    (M_c-M^0_c)D_c\Big) \text{.}
\end{equation}
We can compare the growth with mechanical feedback and the growth without feedback, or purely deformation-led growth, by expressing the growth equation with mechanical feedback in the principal deformation directions ($v_d^i$).
\begin{equation}
    \lambda^i_f = v^{iT}_d\Big[ \kappa(1+\gamma) (M_c-M^0_c)-\frac{\eta}{2} \Big(D_c(M_c-M^0_c) +
    (M_c-M^0_c)D_c\Big) \Big]v^{i}_d
\end{equation}
gives the measure of growth in the direction of principal deformation after the reorganisation of walls by mechanical feedback. We can then measure the change in growth due to stiffness modulation or the relative stiffness modulation due to mechanical feedback by looking at the ratio
\begin{equation}
    \tilde{E}_i = \frac{\lambda^i_d - \lambda^i_f}{\lambda^i_d} \text{.}
\end{equation}
This provides us the measure of growth reorganisation in shape matrix of cells due to the introduction of feedback. As these growth reorganisations are caused by the modulation of stiffness on walls by the CMT dynamics, $\tilde{E}_i$ allows us to gauge the total changes in stiffness caused by the mechanical feedback.\\
By plotting the relative stiffness modulation of cells averaged over the primordial development, the impact of the mechanical feedback is even more striking. Fig.~\ref{fig:stiffnessReorganisationn8} and Fig.~\ref{fig:stiffnessReorganisationn0} show the reorganisation of stiffness by mechanical feedback. The positive values of $\tilde{E}_i$ indicate the strengthening of walls or increase of stiffness, while the negative values represent the loosening of walls. For the case with feedback, the stiffness is increased by up to a factor of $4$ in the boundary cells as compared to the case without.

\newpage
\section{Cell division}

\newpage
\section{Changing bending stiffness}

\newpage
\section{Principal stresses on the cells}

\newpage
\section{Gaussian curvature around the boundary}
\begin{figure}[hbt!]
\centering
	\includegraphics[width=0.8\linewidth]{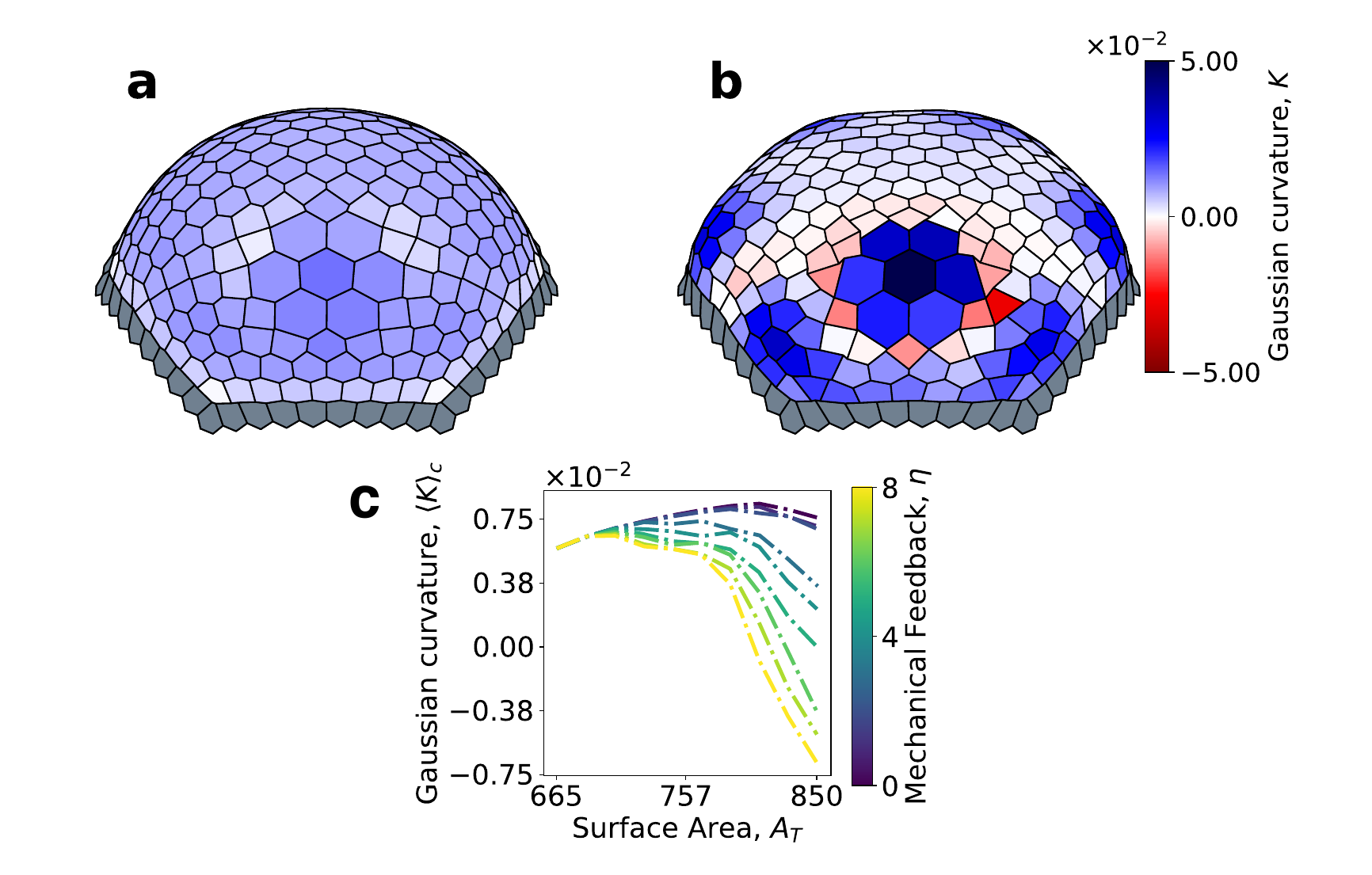}
    \caption{Gaussian curvature of the boundary cells become increasingly negative with mechanical feedback. (a) and (b) show the Gaussian curvature overlaid on the cells for the case of no feedback $\eta = 0$ and with feedback $\eta = 8$ respectively. (c) shows the Gaussian curvature averaged over the boundary cells as a function of tissue surface area for increasing mechanical feedback.}
    \label{fig:gaussiancurvature}
\end{figure}

The saddle shape of the boundary is dependent on the mechanical feedback. With high feedback, the boundary becomes increasingly saddle-like as shown by the negative curvatures arising in the region around primordia in Fig.~\ref{fig:gaussiancurvature}.


\section{Shape of the cells in the tissue}
\begin{figure}[hbt!]
\centering
	\includegraphics[width=0.4\linewidth]{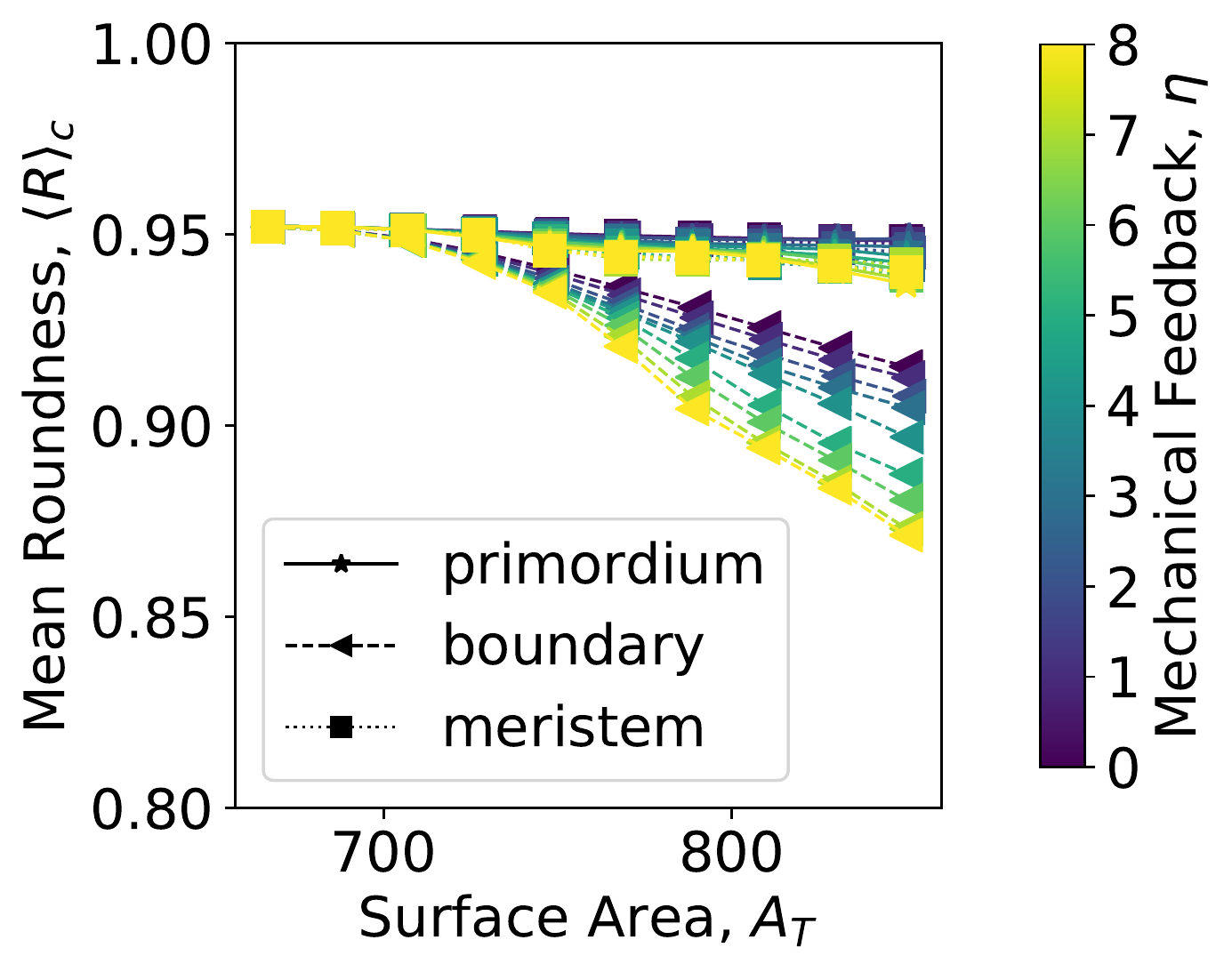}
    \caption{The boundary cells become increasingly anisotropic in shape with mechanical feedback, while the meristem and the primordial cells remain the same.}
    \label{fig:cellroundness}
\end{figure}

To measure the anisotropic shapes of the cells, we use the roundness shape measure for the cells. It is defined as 
\begin{equation}
    R = \frac{2(\pi A_p)^{1/2}}{C_p}
\end{equation}
where, $R$ is the roundness of a cell, $A_p$ is the area of a cell and $C_P$ is the perimeter of a cell. This measure is equal to $1$ for a circle and is increasingly less than $1$ for more and more anisotropic shapes.\\
We observe that the boundary cells become anisotropic with feedback while the primordial and meristematic cells are unchanged (Fig.~\ref{fig:cellroundness}).

\newpage
\section{Feedback model and compressive stress}
The feedback term introduced in Eq.~\ref{eq:restShapeLockhartGrowthFeedback} has the same response to compressive stress as to the expansive stress. The wall reinforcement in the direction of higher magnitude of stress is independent of the stress being positive or negative. 
\newpage

\section{Poisson ratio}
Including Poisson ratio in the simulation, we found that our results are robust under varying the ratio as expected. 

\newpage

\section{Second moment of area plotted for distorted cells}
The shape matrix for cell defined by second moment of area is a robust measure for cell shape. The simple shrinking of walls does not have significant impact on the shape measure as it does not change the areal distribution strongly. This is a major advantage for using cell based vertex model than the wall based one, which can be susceptible to wall length changes. 

\end{document}